%% file: euler.tex
\documentclass[]{llncs}
\usepackage{amssymb}
\usepackage{amsfonts}
\usepackage{mathtools}
\usepackage{calc}
\usepackage{multicol}
\usepackage{multirow}
\usepackage{hyperref}
\usepackage{graphicx}
\usepackage{needspace}

\usepackage{amsmath}

\usepackage{stmaryrd}
\usepackage{xspace}

\usepackage{geometry}
\usepackage{tikz}
\usepackage{pgfplots}
\usetikzlibrary{decorations.markings}
\usetikzlibrary{matrix,backgrounds,folding}
\usetikzlibrary{shapes.geometric}
\pgfdeclarelayer{edgelayer}
\pgfdeclarelayer{nodelayer}
\pgfsetlayers{background,edgelayer,nodelayer,main}
\tikzstyle{every picture}=[baseline=-0.25em]
\tikzstyle{none}=[inner sep=0mm]
\tikzstyle{zxnode}=[shape=circle, minimum width=.25cm, inner sep=0.5pt, font=\footnotesize, draw=black]
\tikzstyle{gn}=[zxnode ,fill=green]
\tikzstyle{rn}=[zxnode ,fill=red]
\tikzstyle{H box}=[rectangle,fill=yellow,draw=black,xscale=1,yscale=1,font=\footnotesize,inner sep=1.2pt,minimum width=0.15cm,minimum height=0.15cm]
\tikzstyle{ug}=[regular polygon, regular polygon sides=3, fill=red,draw=black,inner sep = 0pt,minimum width=1em]
\tikzstyle{black dot}=[inner sep=0.7mm,minimum width=0pt,minimum height=0pt,fill=black,draw=black,shape=circle]
\tikzstyle{dot}=[black dot]
\tikzstyle{white dot}=[dot,fill=white]
\tikzstyle{zwcross}=[diamond, draw, fill=gray, minimum width=0em, inner sep=1.5pt]
\tikzstyle{arrow}=[decoration={markings,mark=at position 1 with
    {\arrow[scale=1.2,>=stealth]{>}}},postaction={decorate}]

\tikzstyle{st}=[star,star points = 5, fill=white,draw=black,inner sep = 1.2pt,line width=1.2pt]
\tikzstyle{gnlabel}=[rounded corners=0.2em,fill=green!20,inner sep=0.1em,font=\scriptsize, anchor=west, xshift=-0.2em, yshift=0,opacity=1]

\pgfdeclarelayer{edgelayer}
\pgfdeclarelayer{nodelayer}
\pgfsetlayers{background,edgelayer,nodelayer,main}
\tikzstyle{none}=[inner sep=0mm]
\tikzstyle{every loop}=[]

\def\fig{}

\newcommand{\callrule}[2]{\hyperlink{r:#1}{\textnormal{(#2)}}\xspace}

\newcommand{\soo}{\callrule{rules}{S}}
\newcommand{\stt}{\callrule{rules}{I}}
\newcommand{\sttr}{\callrule{rules}{I$_r$}}
\newcommand{\sttg}{\callrule{rules}{I$_g$}}
\newcommand{\e}{\callrule{rules}{E}}
\newcommand{\bo}{\callrule{rules}{CP}}
\newcommand{\bt}{\callrule{rules}{B}}
\newcommand{\euler}{\callrule{rules}{EU}}
\newcommand{\eu}{\callrule{rules}{HD}}
\newcommand{\h}{\callrule{rules}{H}}

\newcommand{\eulerp}{\callrule{rules2}{EU'}}
\newcommand{\iv}{\callrule{rules2}{IV}}

\newcommand{\supp}{\callrule{old-rules}{SUP}}
\newcommand{\com}{\callrule{old-rules}{C}}
\newcommand{\bw}{\callrule{old-rules}{BW}}
\newcommand{\kt}{\callrule{old-rules}{K}}
\newcommand{\add}{\callrule{old-rules}{A}}
\newcommand{\ivp}{\callrule{old-rules}{IV'}}
\newcommand{\zo}{\callrule{old-rules}{ZO}}

\newcommand{\comp}[1]{\frac{\pi}{#1}\textnormal{C}\xspace}

\newcommand{\eq}[2][~~]{
#1
\underset{\substack{#2}}{=}
#1
}

\newcommand{\interp}[1]{\left\llbracket #1 \right\rrbracket}

\newcommand{\frag}[1]{$\frac{\pi}{#1}$-frag\-ment}
\newcommand{\titlerule}[1]{\noindent
\begin{center}
\raisebox{0.5ex}{\rule{(\textwidth-\widthof{#1})/2}{0.5pt}}#1\raisebox{0.5ex}{\rule{(\textwidth-\widthof{#1})/2}{0.5pt}}
\end{center}}
\newcommand{\annoted}[3]{{\scriptstyle #1}\left\lbrace\mathrlap{\phantom{#3}}\right.\overbrace{#3}^{#2}}

\newcommand{\ket}[1]{\ensuremath{\left|  #1 \right\rangle}}

\def \zx {\textnormal{ZX}\xspace}

\renewcommand{\cos}[1]{\operatorname{cos}\left(#1\right)}
\renewcommand{\sin}[1]{\operatorname{sin}\left(#1\right)}
\newcommand{\atan}[1]{\operatorname{arctan}\left(#1\right)}

\renewcommand{\tan}[1]{\operatorname{tan}\left(#1\right)}

\newcommand{\Var}{\operatorname{Var}}

\newcommand{\scalar}[1]{\begin{tikzpicture}
	\begin{pgfonlayer}{nodelayer}
		\node [style=none] (0) at (1.75, 0.25) {};
		\node [style=none] (1) at (1.5, -0) {#1};
		\node [style=none] (2) at (1.25, -0.25) {};
		\node [style=none] (3) at (1.25, 0.25) {};
		\node [style=none] (4) at (1.75, -0.25) {};
	\end{pgfonlayer}
	\begin{pgfonlayer}{edgelayer}
		\draw [style=none] (2.center) to (3.center);
		\draw [style=none] (3.center) to (0.center);
		\draw [style=none] (0.center) to (4.center);
		\draw [style=none] (4.center) to (2.center);
	\end{pgfonlayer}
\end{tikzpicture}}

\allowdisplaybreaks

\title{A Near-Optimal Axiomatisation of ZX-Calculus for Pure Qubit Quantum Mechanics}

\author{
Renaud Vilmart
\institute{Universit\'e de Lorraine, CNRS, Inria, LORIA, F 54000 Nancy, France}\\
\email{renaud.vilmart@loria.fr}
}
\begin{document}

\maketitle

\begin{abstract}
Recent developments in the ZX-Calculus have resulted in complete axiomatisations first for an approximately universal restriction of the language, and then for the whole language. The main drawbacks were that the axioms that were added to achieve completeness were numerous, tedious to manipulate and lacked a physical interpretation.

We present in this paper two complete axiomatisations for the general ZX-Calculus, that we believe are optimal, in that all their equations are necessary and moreover have a nice physical interpretation.
\end{abstract}

\section{Introduction}

The ZX-Calculus is a powerful graphical language for quantum computing and reasoning \cite{interacting}. The objects manipulated are open graphs, also called diagrams, that represent quantum evolutions through the standard interpretation. One of the most important features of the language is that the graphs can be considered unoriented, that is, any two isomorphic graphs will yield the same result. Isomorphism between diagrams are not the only transformations that preserve the interpretation though, so the ZX-Calculus comes equipped with a set of axioms: transformations between diagrams that, when applied locally, preserve the interpretation.

The language is universal: any $2^n\times2^m$ matrix can be represented by a ZX-diagram with respect to the standard interpretation. Hence, it has already been used in numerous applications \cite{picturing-qp}, ranging from measure\-ment-based quantum computing \cite{duncan2013mbqc,mbqc,horsman2011quantum} and quantum codes \cite{de2017zx,chancellor2016coherent,verifying-color-code,duncan2014verifying}, to protocols \cite{MSc.Hillebrand} and foundations \cite{toy-model-graph,duncan2016hopf}. The language itself can be manipulated through tools such as Quantomatic \cite{quanto} or PYZX \cite{pyzx}.

A broader use of the ZX-Calculus was limited though, because of a question that remained open for a while: completeness. The language would be complete if, for any two diagrams that represent the same quantum evolution, they could be transformed into one another by mere application of the axioms. The question has been answered for gradually more expressive restrictions of the language. In 2014, complete axiomatisations were provided for the stabiliser \cite{pi_2-complete} and the real stabiliser \cite{pivoting}, then for the one-qubit Clifford+T case \cite{pi_4-single-qubit}. However, none of these restrictions are approximately universal. The first complete axiomatisation for an approximately universal restriction -- the many-qubit Clifford+T -- was recently provided \cite{JPV}, and soon followed two complete axiomatisations for the general -- universal -- ZX-Calculus \cite{HNW,JPV-universal}.

Up to the one-qubit Clifford+T case, all the axioms provided were natural and had a relevant interpretation, however, the axiomatisations for (approximately) universal ZX-Calculus introduced rules that are hard to manipulate, mainly because of their size, and that moreover can not be naturally justified.

We give in this paper a simpler axiomatisation of the general ZX-Calculus, and prove that it is complete for the general ZX-Calculus. It is basically composed of the axioms that make the Clifford -- or stabiliser -- fragment complete, and of an additional axiom, denoted \euler:
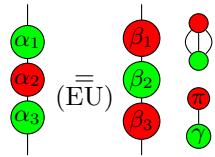
\[
\input{./figures/Euler.tikz}
\]
with a side condition that links the angles on the right to those on the left. In ZX-Calculus, the green node with angle $\alpha$ represents a rotation of angle $\alpha$ around the Z axis (denoted $R_Z(\alpha)$), and the red one a rotation around the orthogonal axis, X (denoted $R_X(\alpha)$). This axiom, which is an application of the Euler angles, essentially gives a normal form for one-qubit unitaries, as a sequences of rotations around the axes X, Z and X again.
This equality between diagrams has been used in \cite{incompleteness} to prove that the then version of ZX-Calculus was not complete, and is part of the axiomatisation of \cite{2-qubits-zx}.

To prove that the new axiomatisation is complete, we simply derive the rules of the former axiomatisation \cite{JPV}. However, since all the power of ``beyond-Clifford'' is contained in the rule \euler, we will end up using it a lot, which would cause a lot of side computation, for the angles on one side of the rule are not defined from the others in a linear fashion. So to avoid having to go through all this tedious process, we use another kind of normal form for ZX-diagrams, which is the graphical version of the singular-value decomposition of a matrix. Hence, instead of showing that a sound equation is derivable, we will show that we can transform the diagrams on both sides into a particular form, which is essentially unique.

We also provide a second axiomatisation, which is not very far from the other. Indeed, in the first, we may notice a rule \eu that we call \emph{the Euler decomposition of Hadamard}, which essentially gives the unitary normal form of the Hadamard gate. The second axiomatisation replaces the rules \eu and \euler by another single rule that unifies them.

In Section \ref{sec:zx}, we formally introduce the language ZX-Calculus, as well as the two aforementioned axiomatisations, and we discuss their minimality. In Section \ref{sec:clifford}, we recover a known complete axiomatisation for the Clifford fragment, hence directly giving us access to already proven lemmas from it. In Section \ref{sec:SVD}, we introduce the singular-value decompositions of cycle-free $0\to1$ and $1\to1$ ZX-diagrams, and show they are essentially unique. Finally, in Section \ref{sec:universal}, we use these decompositions to show the completeness of the axiomatisations for the Clifford+T and for the unrestricted ZX-Calculus.

\section{ZX-Calculus}
\label{sec:zx}

In this section, we introduce the ZX-diagrams together with a new simple axiomatisation that we prove complete in the following sections. The definition of the ZX-diagrams and their interpretation is standard.

\subsection{Diagrams and standard interpretation}

A ZX-diagram $D:k\to l$ with $k$ inputs and $l$ outputs is generated by:
\begin{center}
\bgroup
\def\arraystretch{2.5}
{\begin{tabular}{|cc|cc|}
\hline
$R_Z^{(n,m)}(\alpha):n\to m$ & 
\input{./figures/gn-alpha.tikz}
 & $R_X^{(n,m)}(\alpha):n\to m$ & 
\input{./figures/rn-alpha.tikz}
\\[4ex]\hline
$H:1\to 1$ & 
\input{./figures/Hadamard.tikz}
 & $e:0\to 0$ & 
\input{./figures/empty-diagram.tikz}
\\\hline
$\mathbb{I}:1\to 1$ & 
\input{./figures/single-line.tikz}
 & $\sigma:2\to 2$ & 
\input{./figures/crossing.tikz}
\\\hline
$\epsilon:2\to 0$ & 
\input{./figures/cup.tikz}
 & $\eta:0\to 2$ & 
\input{./figures/caps.tikz}
\\\hline
\end{tabular}}
\egroup\\
where $n,m\in \mathbb{N}$, $\alpha \in \mathbb{R}$, and the generator $e$ is the empty diagram.
\end{center}
and the two compositions:
\begin{itemize}
\item Spacial Composition: for any $D_1:a\to b$ and $D_2:c\to d$, $D_1\otimes D_2:a+c\to b+d$ consists in placing $D_1$ and $D_2$ side by side, $D_2$ on the right of $D_1$.
\item Sequential Composition: for any $D_1:a\to b$ and $D_2:b\to c$, $D_2\circ D_1:a\to c$ consists in placing $D_1$ on the top of $D_2$, connecting the outputs of $D_1$ to the inputs of $D_2$.
\end{itemize}

The standard interpretation of the ZX-diagrams associates to any diagram $D:n\to m$ a linear map $\interp{D}:\mathbb{C}^{2^n}\to\mathbb{C}^{2^m}$ inductively defined as follows:\\
\begin{minipage}{\columnwidth}
\titlerule{$\interp{.}$}
\[ \interp{D_1\otimes D_2}:=\interp{D_1}\otimes\interp{D_2} \qquad 
\interp{D_2\circ D_1}:=\interp{D_2}\circ\interp{D_1}\qquad\interp{
\input{./figures/empty-diagram.tikz}
~}:=\begin{pmatrix}
1
\end{pmatrix} \qquad
\interp{~
\input{./figures/single-line.tikz}
~~}:= \begin{pmatrix}
1 & 0 \\ 0 & 1\end{pmatrix}\]
\end{minipage}
\[\interp{~
\input{./figures/Hadamard.tikz}
~}:= \frac{1}{\sqrt{2}}\begin{pmatrix}1 & 1\\1 & -1\end{pmatrix}\qquad
\interp{
\input{./figures/crossing.tikz}
}:= \begin{pmatrix}
1&0&0&0\\
0&0&1&0\\
0&1&0&0\\
0&0&0&1
\end{pmatrix} \qquad
\interp{\raisebox{-0.35em}{$
\input{./figures/caps.tikz}
$}}:= \begin{pmatrix}
1\\0\\0\\1
\end{pmatrix}\qquad
\interp{\raisebox{-0.25em}{$
\input{./figures/cup.tikz}
$}}:= \begin{pmatrix}
1&0&0&1
\end{pmatrix}\]
\[
\interp{\begin{tikzpicture}
	\begin{pgfonlayer}{nodelayer}
		\node [style=gn] (0) at (0, -0) {$\alpha$};
	\end{pgfonlayer}
\end{tikzpicture}}:=\begin{pmatrix}1+e^{i\alpha}\end{pmatrix} \qquad
\interp{
\input{./figures/gn-alpha.tikz}
}:=
\annoted{2^m}{2^n}{\begin{pmatrix}
  1 & 0 & \cdots & 0 & 0 \\
  0 & 0 & \cdots & 0 & 0 \\
  \vdots & \vdots & \ddots & \vdots & \vdots \\
  0 & 0 & \cdots & 0 & 0 \\
  0 & 0 & \cdots & 0 & e^{i\alpha}
 \end{pmatrix}}
~~\begin{pmatrix}n+m>0\end{pmatrix} 
\]
For any $n,m\geq 0$ and $\alpha\in\mathbb{R}$:\\
\begin{minipage}{\columnwidth}
\[\scalebox{0.9}{$\interp{
\input{./figures/rn-alpha.tikz}
}=\interp{~
\input{./figures/Hadamard.tikz}
~}^{\otimes m}\circ \interp{
\input{./figures/gn-alpha.tikz}
}\circ \interp{~
\input{./figures/Hadamard.tikz}
~}^{\otimes n}$}\] \\
$\left(\text{where }M^{\otimes 0}=\begin{pmatrix}1\end{pmatrix}\text{ and }M^{\otimes k}=M\otimes M^{\otimes k-1}\text{ for any }k\in \mathbb{N}^*\right)$.\\
\rule{\columnwidth}{0.5pt}
\end{minipage}\\

To simplify, the red and green nodes will be represented empty when holding a 0 angle:
\[ \scalebox{0.9}{
\input{./figures/gn-empty-is-gn-zero.tikz}
} \qquad\text{and}\qquad \scalebox{0.9}{
\input{./figures/rn-empty-is-rn-zero.tikz}
} \]

ZX-Diagrams are universal:
\[\forall A\in \mathbb{C}^{2^n}\times\mathbb{C}^{2^m},~~\exists D:n\to m,~~ \interp{D}=A\]

However, it is customary to restrict the language to a countable or finite set of angles. Some of these restrictions, or fragments, correspond to well-known restrictions of quantum computing: The \frag2 -- the restriction where all the angles are multiples of $\frac{\pi}{2}$ -- corresponds to Clifford; while the \frag4 corresponds to Clifford+T. In the following, we may refer to the \frag2 using the term Clifford, and similarly for the \frag4.

\subsection{Calculus}

\begin{figure*}[!htb]
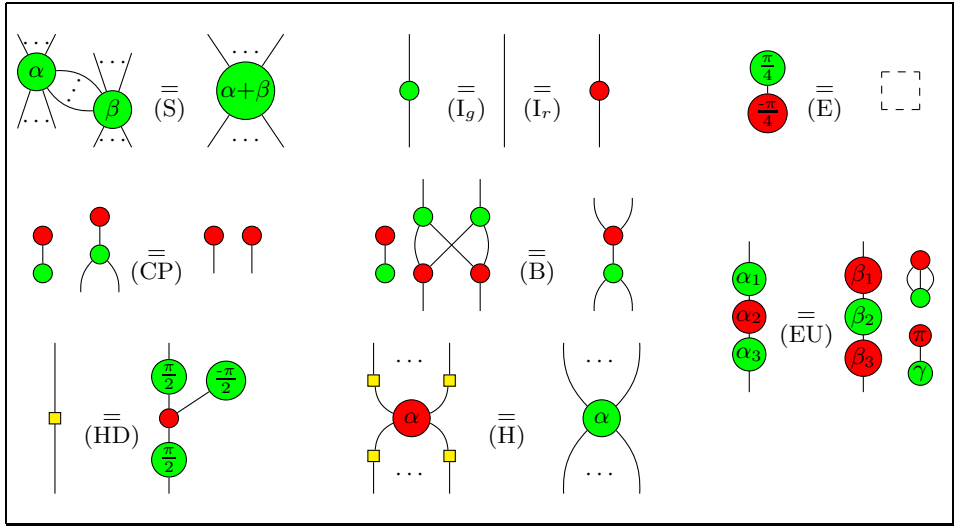

 \centering
 \hypertarget{r:rules}{}
 \scalebox{1}{\begin{tabular}{|c@{$\qquad\quad~$}c@{$\qquad\quad~$}c@{~}|}
   \hline
   && \\
   
\input{./figures/spider-1.tikz}
&
\input{./figures/s2-green-red.tikz}
&
\input{./figures/bicolor_pi_4_eq_empty.tikz}
\\
   && \\
   
\input{./figures/b1s.tikz}
&
\input{./figures/b2s.tikz}
&\multirow{3}{*}{
\input{./figures/Euler.tikz}
}\\
   && \\
   
\input{./figures/euler-decomp-scalar-free.tikz}
&
\input{./figures/h2.tikz}
&\\
   && \\
   \hline
  \end{tabular}}
 \caption[]{Set of rules \zx for the ZX-Calculus with scalars. The right-hand side of (E) is an empty diagram. (...) denote zero or more wires, while (\protect\rotatebox{45}{\raisebox{-0.4em}{$\cdots$}}) denote one or more wires. In rule (EU), $\beta_1,\beta_2,\beta_3$ and $\gamma$ can be determined as follows:
$x^+:=\frac{\alpha_1+\alpha_3}{2}$, $x^-:=x^+-\alpha_3$, $z := \cos{\frac{\alpha_2}{2}}\cos{x^+}+i\sin{\frac{\alpha_2}{2}}\cos{x^-}$ and $
z' := \cos{\frac{\alpha_2}{2}}\sin{x^+}-i\sin{\frac{\alpha_2}{2}}\sin{x^-}$, then
$\beta_1 = \arg z + \arg z',
\beta_2 = 2\arg\left(i+\left|\frac{z}{z'}\right|\right),
\beta_3 = \arg z - \arg z',
\gamma = x^+-\arg(z)+\frac{\alpha_2-\beta_2}{2}$
where by convention $\arg(0):=0$ and $z'=0\implies \beta_2=0$.
  }
 \label{fig:ZX_rules}
\end{figure*}

The diagrammatic representation of a matrix is not unique in the ZX-Calculus. As a consequence the language comes with a set of axioms. Additionally to the axioms of the language described in Figure \ref{fig:ZX_rules}, one can:
\begin{itemize}
\item bend any wire  of a ZX-diagram at will, without changing its semantics. This paradigm -- the so-called \textbf{Only Connectivity Matters} -- can be derived from the following axioms:
\[{
\input{./figures/bent-wire.tikz}
}\]
\[{
\input{./figures/bent-wire-2.tikz}
}\]
\item apply the axioms to sub-diagrams. If $\zx\vdash D_1=D_2$ then, for any diagram $D$ with the appropriate number of inputs and outputs:
\begin{itemize}
\item $\zx\vdash D_1\circ D = D_2\circ D$
\item $\zx\vdash D\circ D_1 = D\circ D_2$
\item $\zx\vdash D_1\otimes D = D_2\otimes D$
\item $\zx\vdash D\otimes D_1 = D\otimes D_2$
\end{itemize}
where $\zx\vdash D_1 = D_2$ means that $D_1$ can be transformed into $D_2$ using the axioms of the ZX-Calculus. 
\end{itemize}

All the axioms of Figure \ref{fig:ZX_rules}, but \euler,  are standard in the ZX-calculus. Roughly speaking: \soo and \stt correspond to the axiomatisation of orthonormal basis \cite{CPV12}, each color being associated with an orthonormal basis; \bo and \bt capture the fact that the two bases are strongly complementary \cite{interacting};  \h  means that Hadamard can be used to exchange the colours and \eu means that Hadamard can be decomposed using $\frac \pi 2$-rotations \cite{euler-decomp}; \e states that some particular scalars (ZX-diagram with no input/output) can vanish, which means that their interpretation is one \cite{cyclo}. In the following we investigate the properties of \euler.

\subsection{The Euler Angles}
\label{sec:euler-angles}

The rule \euler is really all about unitaries. Indeed, we have the following result:

\begin{proposition}
\label{prop:unitary-uniqueness}
Any one-qubit unitary can be decomposed as $e^{i\gamma}R_Z(\alpha_3)R_X(\alpha_2)R_Z(\alpha_1)$, which can be represented in ZX as:
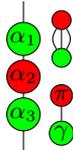
\[
\input{./figures/one-qubit-unitary.tikz}
\]
If the unitary is not diagonal or anti-diagonal (i.e.~if $\alpha_2\neq0\bmod\pi$), then this decomposition can be made unique if we impose $\alpha_1\in[0,\pi)$
\end{proposition}

In 1775, Euler proved what is now called Euler's rotation theorem \cite{euler-angles}, stating that there are several ways to decompose a rotation into several rotations around elementary axes. In quantum mechanics, a consequence is that any unitary operator on one qubit can be seen as either a composition of 
rotations around Z, X, Z; or around X, Z, X.
 On the one hand, the rule \eu says -- in a distorted, ZX-style way -- that the Hadamard gate can be decomposed as a series of rotations, while on the other hand, the rule \euler gives the equality between two different decompositions of the same unitary:
\[
\input{./figures/Euler.tikz}

\quad\text{where}\quad\left\lbrace\begin{array}{l}
x^+:=\frac{\alpha_1+\alpha_3}{2}\qquad x^-:=x^+-\alpha_3\\
z := \cos{\frac{\alpha_2}{2}}\cos{x^+}+i\sin{\frac{\alpha_2}{2}}\cos{x^-}\\
z' := \cos{\frac{\alpha_2}{2}}\sin{x^+}-i\sin{\frac{\alpha_2}{2}}\sin{x^-}\\
\beta_1 = \arg z + \arg z\\
\beta_2 = 2\arg\left(i+\left|\frac{z}{z'}\right|\right)\\
\beta_3 = \arg z - \arg z'\\
\gamma = x^+-\arg(z)+\frac{\alpha_2-\beta_2}{2}
\end{array}\right.
\] 
The angles $\beta_i$ and $\gamma$ seem to not always be defined. Indeed, $\arg$ is not defined in $0$, and $\beta_2$ is not defined when $z'=0$. By convention, we decide that $\arg(0)=0$ and that $\beta_2=0$ when $z'=0$.

The first proof of incompleteness \cite{incompleteness} relied on an euler decomposition, but adding it to the set of ZX axioms has been avoided for a while because of its non-linearity. However, a non-linear axiom is necessary to get the completeness for the general ZX-Calculus \cite{JPV-universal}. And so, it has been used in \cite{2-qubits-zx} to prove the completeness of the 2-qubit \frag4 of the ZX-Calculus. The rule \euler is actually much more powerful than this, for, as we will prove in the following:

\begin{theorem}
\label{thm:completeness}
The ZX-Calculus -- with axioms in Figure \ref{fig:ZX_rules} -- is complete for pure qubit quantum mechanics. For any two diagrams $D_1$ and $D_2$ of the ZX-Calculus:
\[\interp{D_1}=\interp{D_2} \iff \zx\vdash D_1=D_2 \]
\end{theorem}

\subsection{On Minimality}

We call an axiomatisation minimal when there is no redundancy in the axioms. Particularly, we want a proof that none of the axioms are derivable from the others. We conjecture that all the axioms in Figure \ref{fig:ZX_rules} are necessary. Indeed, in \cite{minimal-stabiliser-zx}, nearly all the rules for Clifford -- i.e.~all of the axioms in Figure \ref{fig:ZX_rules} except \e and \euler -- are proven to be necessary, and all arguments stand here:
\begin{itemize}
\item \soo: It is the only axiom that can transform a node of degree four or higher into a diagram containing lower-degree nodes
\item \sttg or \sttr: These are the only two axioms that can transform a diagram with nodes connected to a boundary to a node-free diagram
\item \bo: It is the only axiom that can transform a diagram with two connected outputs into one with two disconnected outputs
\item \eu: The necessity of this axiom requires a non-trivial interpretation given in \cite{euler-decomp,pivoting}, and given again in the Appendix at page \pageref{prf:hd-necessity}.
\item \h: It is the only axiom that matches red nodes with 4+ degree to green nodes of the same degree
\end{itemize}
However, \e and \euler can also be proven to be necessary:
\begin{itemize}
\item \e: It is the only axiom that can transform a non-empty diagram into an empty one
\item \euler: It is the only non-linear axiom
\end{itemize}
In a nutshell, all the axioms are proven to be necessary, except \bt and one of the \stt.

\begin{figure*}[!htb]
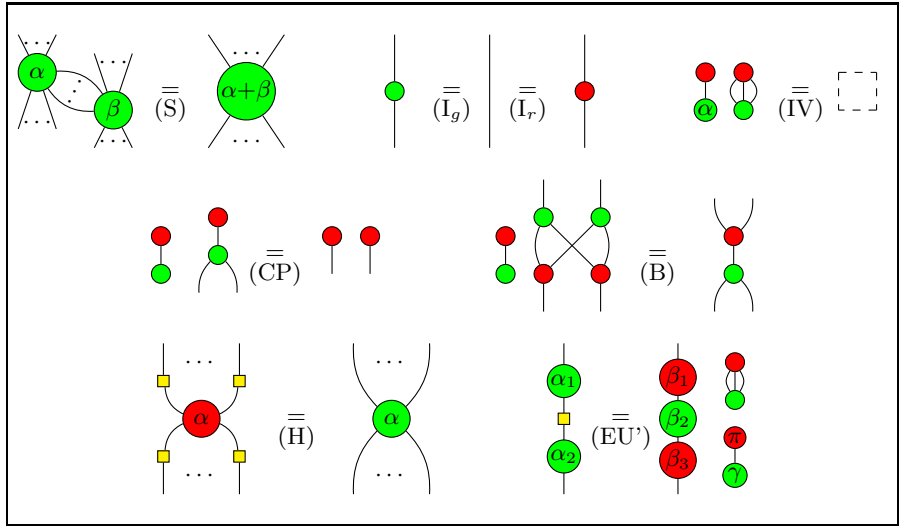

 \centering
 \hypertarget{r:rules2}{}
 \scalebox{1}{\begin{tabular}{|c|}
   \hline\\
   
\input{./figures/spider-1.tikz}
$\qquad\quad~$
\input{./figures/s2-green-red.tikz}
$\qquad\quad~$
\input{./figures/inverse-param.tikz}
~~\\\\
   
\input{./figures/b1s.tikz}
$\qquad\qquad$
\input{./figures/b2s.tikz}
\\\\
   
\input{./figures/h2.tikz}
$\qquad\qquad$
\input{./figures/Euler-Had.tikz}
\\\\
   \hline
  \end{tabular}}
 \caption[]{Set of rules \zx' for the ZX-Calculus with scalars. The right-hand side of (E) is an empty diagram. (...) denote zero or more wires, while (\protect\rotatebox{45}{\raisebox{-0.4em}{$\cdots$}}) denote one or more wires. In rule (EU'), $\beta_1,\beta_2,\beta_3$ and $\gamma$ can be determined as follows:
$x^+:=\frac{\alpha_1+\alpha_2}{2}$, $x^-:=x^+-\alpha_2$, $z := -\sin{x^+}+i\cos{x^-}$ and $
z' := \cos{x^+}-i\sin{x^-}$, then
$\beta_1 = \arg z + \arg z',
\beta_2 = 2\arg\left(i+\left|\frac{z}{z'}\right|\right),
\beta_3 = \arg z - \arg z',
\gamma = x^+-\arg(z)+\frac{\pi-\beta_2}{2}$
where by convention $\arg(0):=0$ and $z'=0\implies \beta_2=0$.
  }
 \label{fig:ZX_rules2}
\end{figure*}

Another aspect of minimality, is whether a rule can be made ``simpler'' thanks to the others, according to some measure, be it arbitrary or well-defined. In the previous axiomatisation, we have two rules that are closely related to how unitaries can be decomposed: \eu and \euler. It so happens that we can fuse them into one, of the same size as \euler, and doing so allows us to simplify the scalar rule:

\begin{theorem}
\label{thm:equivalence}~\\
$\left(~~
\input{./figures/euler-decomp-scalar-free.tikz}
,~~
\input{./figures/Euler.tikz}
,~~
\input{./figures/bicolor_pi_4_eq_empty.tikz}
~\right)$ can be replaced by\\
$\left(~~
\input{./figures/Euler-Had.tikz}
,~~
\input{./figures/inverse-param.tikz}
~\right)$
$
\text{where}\quad\left\lbrace\begin{array}{l}
x^+:=\frac{\alpha_1+\alpha_2}{2}\qquad x^-:=x^+-\alpha_2\\
z := -\sin{x^+}+i\cos{x^-}\\
z' := \cos{x^+}-i\sin{x^-}\\
\beta_1 = \arg z + \arg z\\
\beta_2 = 2\arg\left(i+\left|\frac{z}{z'}\right|\right)\\
\beta_3 = \arg z - \arg z'\\
\gamma = x^+-\arg(z)+\frac{\pi-\beta_2}{2}
\end{array}\right.$

As a consequence, the axiomatisation given in Figure \ref{fig:ZX_rules2} is complete for universal quantum mechanics.
\end{theorem}

\begin{proof}
The proof, done at the end of the appendix, at page \pageref{prf:equivalence}, consists in showing that all the rules in Figure \ref{fig:ZX_rules} are derivable.
\end{proof}

On the one hand, this new axiomatisation is one axiom shorter, and \eulerp and \iv can be considered simpler than \euler and \e. On the other hand, the axiomatisation in Figure \ref{fig:ZX_rules} has the nice property that it suffices to remove \euler and \e to get a complete axiomatisation for the scalar-free Clifford fragment. Moreover, \euler is arguably more natural, and has already been given for instance in \cite{2-qubits-zx}.

The following of the paper is dedicated to the proof of Theorem \ref{thm:completeness}. Since \cite{HNW,JPV-universal} provided us with two complete axiomatisations for the general ZX-Calculus, all we have to do is prove all the equations used as axioms in either one of these two axiomatisations. As the axiomatisation in \cite{HNW} requires additional generators and more axioms, we will use the axiomatisation of \cite{JPV-universal} as a reference which consists in all the axioms of Figure \ref{fig:ZX_rules}  but \euler, together with the following axioms, we call obsolete, as we are proving in the following that they can be derived using the (EU) rule:

\noindent\begin{minipage}{\columnwidth}
\hypertarget{r:old-rules}{}
\titlerule{~Obsolete ZX-rules~}
\centering

\input{./figures/k2s-label.tikz}
\hspace*{2em}
\input{./figures/former-supp-label.tikz}
\hspace*{2em}
\input{./figures/commutation-of-controls-general-simplified.tikz}

\end{minipage}\\\\~\\
{\centering

\input{./figures/BW-simplified.tikz}
\hspace*{5em}
\input{./figures/add-axiom.tikz}
\\~\\}
\noindent\begin{minipage}{\columnwidth}
\def\fig{zero-proof-4}
\begin{align*}
\input{./figures/\fig/\fig_00.tikz}\eq{\textnormal{\normalsize (Z0)}}\input{./figures/\fig/\fig_05.tikz}\qquad\qquad\qquad
\def\fig{scalar-inverse-proof}\input{./figures/\fig/\fig_00.tikz}\eq{\textnormal{\normalsize (IV')}}\input{./figures/\fig/\fig_07.tikz}
\end{align*}
\rule{\columnwidth}{0.5pt}
\end{minipage}
 
\begin{remark}
The last two equations, \zo and \ivp, are actually derivable from \kt, \supp and the Clifford axiomatisation \cite{cyclo}. However, they are given here, because together with \soo, \stt, \bo, \bt, \eu and \h, they make the Clifford fragment complete, which will be our first milestone.
\end{remark} 

\section{Clifford}

\label{sec:clifford}

As we just said, a first and easy step to do is to show that we can recover the rules that are known to make the language complete for Clifford \cite{minimal-stabiliser-zx}. This will allow us to freely use in the following all the equations of the \frag2 that are sound. We already have most of these rules that make the ZX-calculus complete for Clifford. We only lack two: the \emph{zero} \zo and the \emph{inverse} \ivp rules. A first very well known lemma we will use for both proofs is the Hopf law:

\begin{lemma}[Hopf Law]
\label{lem:hopf}
\def\fig{hopf}
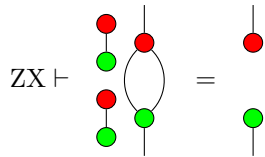
\[\zx\vdash~~\input{./figures/\fig/\fig_00.tikz}\eq{}\input{./figures/\fig/\fig_05.tikz}\]
\end{lemma}

From there, it is fairly easy to recover the inverse rule:

\begin{proposition}
\label{prop:inverse}
The  inverse rule is derivable:
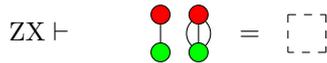
\[\zx\vdash~~\qquad\def\fig{scalar-inverse-proof}\input{./figures/\fig/\fig_00.tikz}\eq{}\input{./figures/\fig/\fig_07.tikz}\]
\end{proposition}

To prove the zero rule, we will use another well known equation, called \emph{$\pi$-commutation}, which is also one of the now obsolete rules.

\begin{proposition}
\label{prop:pi-commutation}
The $\pi$-commutation is derivable:
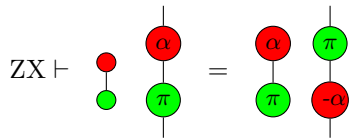
\def\fig{k2s-proof}
\begin{align*}
\zx\vdash ~~\input{./figures/\fig/\fig_00.tikz}
\eq{}\input{./figures/\fig/\fig_03.tikz}
\end{align*}
\end{proposition}

and, with some effort, the rule \zo, which only deals with null diagrams, can be recovered:

\begin{proposition}
\label{prop:zero}
The zero rule is derivable:
\def\fig{zero-proof-4}
\begin{align*}
\zx\vdash~~\input{./figures/\fig/\fig_00.tikz}
\eq{}\input{./figures/\fig/\fig_05.tikz}
\end{align*}
\end{proposition}

As a result:
\begin{theorem}
For any diagrams $D_1$, $D_2$ of the \frag2:
\[\interp{D_1}=\interp{D_2} \iff \zx\vdash D_1 = D_2\]
\end{theorem}

From this first milestone, we get all the sound equations in Clifford, but actually also a bit more. For instance, the following lemmas are known to be derivable from the Clifford axiomatisation (see Appendix):
\begin{multicols}{2}
\begin{lemma}
\label{lem:multiplying-phases}
\[
\input{./figures/multiplying-global-phases.tikz}
\]
\end{lemma}
\begin{lemma}
\label{lem:bicolor-0-alpha}
\def\fig{bicolor-0-alpha}
\[\input{./figures/\fig/\fig_00.tikz}\eq{}\input{./figures/\fig/\fig_07.tikz}\]
\end{lemma}
\end{multicols}

\section{Singular Value Decompositions}

\label{sec:SVD}

The next step is logically to get the completeness for Clifford+T quantum mechanics, i.e. the completeness of the  $\frac \pi 4$-fragment of the ZX-calculus.
Now that we are seeking to prove equations that are out of Clifford, we will begin to use \euler to its full potential. However, we would like, as much as possible, to avoid computing the angles, because, since we work on the problem of completeness, we need to \emph{formally} prove the equality between two diagrams, and hence to formally write what the angles resulting from \euler are, which becomes tedious after a few number of application of the rule.

To simplify this task, instead of showing directly that two diagrams can be turned into one another, we will define a normal form for them, show that it is unique, and show that there is an algorithm to turn them in this normal form..

First, we show another version of the rule \euler with dangling branches:

\begin{lemma}
\label{lem:euler-branch}
\def\fig{Euler-branch}
\[\input{./figures/\fig/\fig_00.tikz}\eq{}\input{./figures/\fig/\fig_03.tikz}\]
where $\beta_1,\beta_2,\beta_3,\gamma$ can be determined as in rule \euler.
\end{lemma}
In a particular case, it implies:

\begin{corollary}
\label{cor:euler-branch}
\def\fig{Euler-branch-cor-2}
\[\input{./figures/\fig/\fig_00.tikz}\eq{}\input{./figures/\fig/\fig_03.tikz}\]
where $\beta_1,\beta_2,\beta_3,\gamma$ can be determined as in rule \euler with $\alpha_2 \leftarrow \frac{\pi}{2}$.
\end{corollary}

We can also derive a kind of inverse operation:

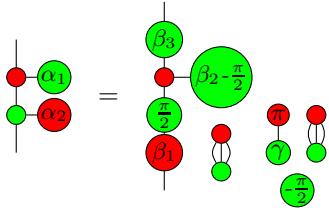
\begin{lemma}
\label{lem:two-branches-to-one}
\def\fig{two-branches-to-one-by-euler-2}
\[\input{./figures/\fig/\fig_00.tikz}
\eq{}\input{./figures/\fig/\fig_06.tikz}\]
where $\beta_1,\beta_2,\beta_3,\gamma$ can be determined as in rule \euler applied with the angles $\alpha_2\leftarrow \alpha_2+\frac{\pi}{2}$ and $\alpha_3\leftarrow\frac{\pi}{2}$.
\end{lemma}

Then, we show that any diagram in the form of the left hand side of \supp -- but with arbitrary angles -- can be transformed in a state with no branching:

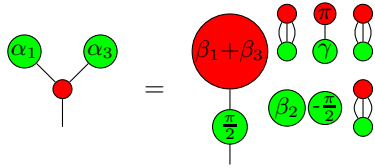
\begin{lemma}
\label{lem:sup-gen}
\def\fig{Euler-branch-to-line-3}
\[\input{./figures/\fig/\fig_00.tikz}
\eq{}\input{./figures/\fig/\fig_03.tikz}\]
where $\beta_1,\beta_2,\beta_3,\gamma$ can be determined as in rule \euler with $\alpha_2 \leftarrow \frac{\pi}{2}$.
\end{lemma}

Now, by specialising the angles to $\alpha$ and $\alpha+\pi$, we shall recover \supp:

\begin{proposition}
\label{prop:supp}
The supplementarity is derivable:
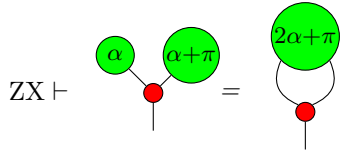
\[\zx\vdash ~~ 
\input{./figures/former-supp.tikz}
\]
\end{proposition}

\begin{remark}
\label{rem:gn-pi_2-0-0-equals-sqrt2-exp-pi_4}
The supplementarity allows us to prove:
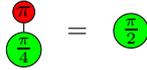
\def\fig{gn-pi_2-0-0-equals-sqrt2-exp-pi_4-proof}
\begin{align*}
\input{./figures/\fig/\fig_00.tikz}
\eq{}\input{./figures/\fig/\fig_06.tikz}
\end{align*}
which, coupled with Lemma \ref{lem:multiplying-phases}
, implies that \def\fig{remark-gamma-pi_2}\input{./figures/\fig/\fig_00.tikz} can be replaced by \input{./figures/\fig/\fig_01.tikz} in the last three lemmas.
\end{remark}

Right now, we have proven all the equations that do not really need a unique normal form. For the rest, we present the singular-value decomposition of a matrix, and introduce it to ZX-diagrams.

\begin{definition}
We call a singular value decomposition (SVD) of a matrix a decomposition of the form 
\[M = U\Sigma V^{\dagger}\]
where $U$ and $V$ are unitary, and $\Sigma$ is diagonal. Notice that $M$ needs not be square (in this case $\Sigma$ has the same dimensions as $M$).
\end{definition}

To justify the use of SVDs, we give some of it interesting properties \cite{SVD}:

\begin{proposition}
\label{prop:SVD-properties}
The SVD $M = U\Sigma V^{\dagger}$ of a matrix $M$ has the following properties:
\begin{itemize}
\item It exists whatever $M$
\item $\Sigma$ can be made unique if we impose that its diagonal entries are decreasing non-negative real numbers
\item $U$ and $V$ are not unique in general, though:
\item If $M$ is square with distinct and non-zero singular values, then $U$ and $V$ are essentially unique:
\[ U\Sigma V^{\dagger} = U'\Sigma V'^{\dagger} \iff \left(\exists d, (U'=Ud) \wedge (V'=Vd)\right)\]
where $d$ is diagonal with diagonal entries some roots of unity.
\end{itemize}
\end{proposition}

Even though the singular-value decomposition is relevant for any diagram, we are only going to give its derivation for a particular family of diagrams:

\begin{definition}
We call a \emph{cycle-free} diagram a diagram composed only of 
\input{./figures/single-line.tikz}
, 
\input{./figures/Hadamard.tikz}
, 
\input{./figures/gn-n-1.tikz}
, 
\input{./figures/rn-n-1.tikz}
 where $n\in\mathbb{N}$ and $\alpha\in\mathbb{R}$.
\end{definition}

\begin{remark}
Some diagrams that do not strictly follow the conditions of the previous definition will still be considered cycle-free if they are equal to a cycle-free diagram by mere application of the ``only topology matters'' paradigm, i.e.~if they are isomorphic to a cycle-free diagram. E.g.:
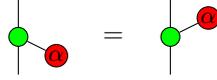
\[
\input{./figures/cycle-free-diagram-example.tikz}
\]
\end{remark}

We can now easily give a normal form for one-qubit states, using the SVD. In this case $\Sigma=s'\ket{0}$, $U$ is a one-qubit unitary, which can be expressed as in Proposition \ref{prop:unitary-uniqueness}, and $V$ is a $0\to 0$ unitary, i.e.~a global phase.

\begin{proposition}[SVD of a State]
\label{prop:state-SVD}
Any cycle-free state $D:0\to1$ can be put in the following forms:
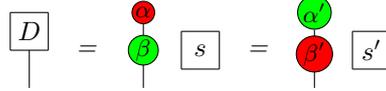
\[
\input{./figures/generic-state.tikz}
\]
where $\beta,\beta'\in[0,\pi)$, and where $s$ and $s'$ are $0\to 0$ diagrams, i.e.~scalars. We call these two forms respectively SVD$_g$ and SVD$_r$.
\end{proposition}

\begin{proposition}[SVDs of states are essentially unique]
\label{prop:state-SVD-uniqueness}
If $D_1=
\input{./figures/generic-state-1.tikz}
~\scalar{$s_1$}$ and $D_2=
\input{./figures/generic-state-2.tikz}
~\scalar{$s_2$}$ are in SVD, and if $\interp{D_1}=\interp{D_2}\neq 0$, then either:
\begin{itemize}
\item $\alpha_1=\alpha_2\bmod 2\pi$ and $\alpha_i=0\bmod \pi$
\item $\alpha_1=\alpha_2$ and $\beta_1=\beta_2$
\end{itemize}
\end{proposition}

We can have basically the same results for $1\to1$ operators:

\begin{proposition}[SVD of a $1\to 1$ diagram]
\label{prop:1-1-diag-SVD}
Any cycle-free diagram $D:1\to 1$ can be written in the forms:
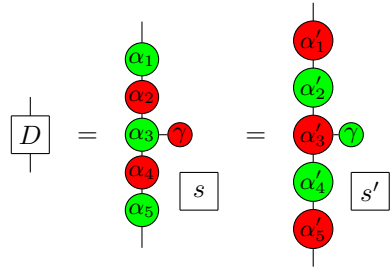
\def\fig{1-1-diagram-decomposition-by-singular-values}
\begin{align*}
\input{./figures/\fig/\fig_00.tikz}
\eq{}\input{./figures/\fig/\fig_01.tikz}
\eq{}\input{./figures/\fig/\fig_04.tikz}
\end{align*}
where $\gamma\in[0,\frac{\pi}{2}]$, and $\alpha_1,\alpha_5,\alpha'_1,\alpha'_5\in[0,\pi)$. We denote the two forms respectively SVD$_g$ and SVD$_r$.
\end{proposition}

\begin{remark}
We gave two conventions for the SVDs of $0\to1$ and  $1\to 1$ diagrams. These two depend on the basis in which we consider the decomposition. SVD$_g$ corresponds to the computational basis, while SVG$_r$ corresponds to the diagonal basis. If $M=U\Sigma V^{\dagger}$ with $\Sigma$ diagonal in the computational basis, $M = (UH)\cdot H\Sigma H \cdot (VH)^{\dagger}$.
\end{remark}

\begin{proposition}[$1\to 1$ SVDs are essentially unique]~\\
\label{prop:1-1-SVD-essentially-unique}
Suppose \def\fig{1-1-diagram-decomposition-by-singular-values}$D_1=\input{./figures/\fig/\fig_01.tikz}$ and $D_2=
\input{./figures/D2-SVD.tikz}
$ are in SVD, and that $\interp{D_1}=\interp{D_2}\neq 0$. Then, either:
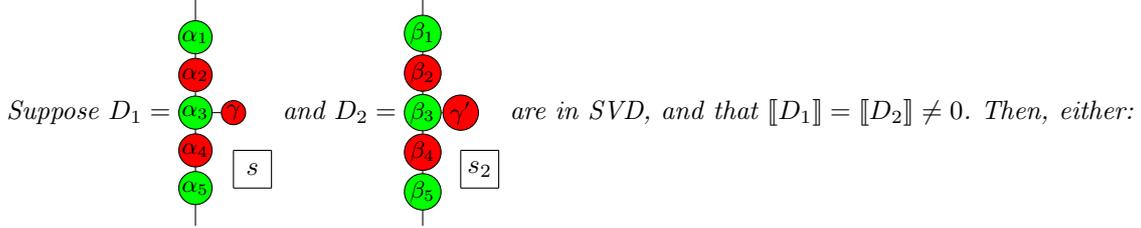
\begin{itemize}
\item $\gamma = \gamma' = 0$
\item $\gamma = \gamma' = \frac{\pi}{2}$
\item $\alpha_i=\beta_i$ and $\gamma=\gamma'$
\end{itemize}
\end{proposition}

\section{Clifford+T and Beyond}

\label{sec:universal}
The point now is to exploit the SVD of ZX-diagrams and their uniqueness. A rule that can directly use these results is \bw:

\begin{proposition}
\label{prop:BW}
\def\fig{BW-simplified-proof}
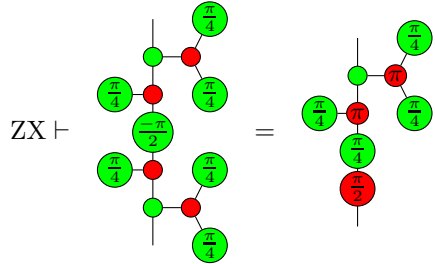
\begin{align*}
\zx\vdash~~\input{./figures/\fig/\fig_00.tikz}
\eq{}\input{./figures/\fig/\fig_01.tikz}
\end{align*}
\end{proposition}

The results on SVDs can not be directly used to prove the equation \com, for its diagrams have 4 inputs/outputs, and have a cycle.
However, the SVDs can be used to prove a first intermediary result:

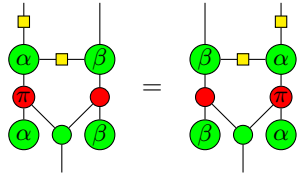
\begin{lemma}
\label{lem:C1-h}
\begin{equation*}
\def\fig{C1-proof}\input{./figures/\fig/\fig_00.tikz}\eq{}\def\fig{C1-proof-2}\input{./figures/\fig/\fig_00.tikz}
\end{equation*}
\end{lemma}
From which we can deduce the equation \com itself:

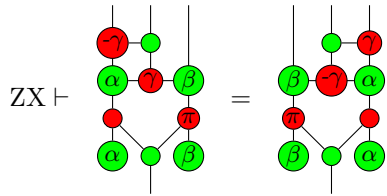
\begin{proposition}
\label{prop:rule-C}
\def\fig{C-proof}
\begin{align*}
\zx\vdash~~\input{./figures/\fig/\fig_00.tikz}
\eq{}\input{./figures/\fig/\fig_08.tikz}
\end{align*}
\end{proposition}

\begin{remark}
From Lemma \ref{lem:C1-h}, \com can be derived using only the Clifford rules. However, the provided proof requires using half angles. Hence, whenever the considered fragment contains all its half angles, the equation in Lemma \ref{lem:C1-h} should be preferred to \com.
\end{remark}

We have derived all the rules necessary for the completeness of the Clifford+T fragment of the ZX-Calculus, which means:

\begin{theorem}
For any diagrams $D_1$, $D_2$ of the \frag4:
\[\interp{D_1}=\interp{D_2} \iff \zx\vdash D_1 = D_2\]
\end{theorem}

Finally, it remains to derive the equation \add. Notice that the diagram on the left hand side contains a cycle, which implies we can not use the results on SVDs. However, the cycle can be easily removed, and we are able to prove:

\begin{proposition}
\label{prop:add-axiom}
\[\zx\vdash 
\input{./figures/add-axiom-no-label.tikz}
\]
\end{proposition}

This last proposition ends the proof of Theorem \ref{thm:completeness}.\qed

\section{Discussion and Further Work}

We have provided two simple but complete axiomatisations of the ZX-Calculus for universal quantum mechanics. By doing so, we have restored intuitiveness -- one of its the first aims -- to the language (at least on the structural level, computing the angles in \euler remains tedious if done formally). This step forward should simplify axiom-related problems such as verification or compilation. 

To simplify the task of proving the derivability of equations, we introduced singular-value decomposition of $0\to1$ and $1\to1$ diagrams, and proved that there exists an algorithm to turn any $0\to1$ and $1\to1$ cycle-free diagram into its SVD form. We did not need SVD form for diagrams \emph{with} cycle, and leave as a further development the extension of the algorithm to arbitrary $0\to1$ and $1\to1$ diagrams, which should be possible by completeness and universality.

We did not need to define the SVD form for larger diagrams either. A problem would arise in ZX, for instance for a diagram with 3 inputs/outputs: do we decompose the diagram as a $0\to3$, or a $1\to2$ diagram and then use the map/state duality? This would result in two completely different decompositions. Still, defining SVDs for diagrams of any arity could prove interesting.

Concerning the result itself, we have proven that, in ZX-Calculus:
\begin{center}
\boxed{\left.
\begin{array}{c}
$many-qubit Clifford completeness$\\
+\\
$completeness for 1-qubit unitaries$
\end{array}
\right\rbrace
 =$ many-qubit completeness$}
\end{center}
This formulation is a bit excessive, since we actually have several rules that operate beyond the Clifford fragment, namely \soo and \h, where the angles can take any value in $\mathbb{R}$ -- and this feature is actually needed for the completeness. Still, since it is not absurd to imagine we can always find similar rules for the considered language, this raises two questions:
\begin{itemize}
\item Is it true for fragments of the ZX-Calculus?\\
The answer in general is no. Indeed, in the case of Clifford+T, the axiomatisation for Clifford is enough to get the 1-qubit completeness \cite{pi_4-single-qubit}. However, it has been proven that rules \supp and \e are necessary \cite{cyclo,supplementarity}. Hence, the previous statement does not stand for Clifford+T.
\item How far from this statement are we in other languages?\\
For instance, we know a complete presentation for the many-qubit Clifford fragment of quantum circuits \cite{clifford-circuits}. Moreover, the rule \euler has an obvious equivalent in circuits, and is the only needed axiom for 1-qubit completeness. So what do we lack to get the universal completeness?
\end{itemize}

\section*{Acknowledgements}

The author acknowledges support from the projects ANR-17-CE25-0009 SoftQPro, ANR-17-CE24-0035 VanQuTe, PIA-GDN/Quantex, and STIC-AmSud 16-STIC-05 FoQCoSS. All diagrams were written with the help of TikZit.

\appendix

\section{Appendix}

\begin{proof}[Prop. \ref{prop:unitary-uniqueness}]~\\
$\bullet$ Existence:\\
Any element of $U(2)$ can be decomposed as:
\[e^{i\varphi/2}\begin{pmatrix}e^{i\psi_0}&0\\0&e^{-i\psi_0}\end{pmatrix}\begin{pmatrix}\cos{\theta}&\sin{\theta}\\-\sin{\theta}&\cos{\theta}\end{pmatrix}\begin{pmatrix}e^{i\psi_1}&0\\0&e^{-i\psi_1}\end{pmatrix}\]
Hence, the existence is given by:
\begin{align*}
\interp{
\input{./figures/one-qubit-unitary.tikz}
} &= 
e^{i(\gamma+\frac{\alpha_2}{2})}\begin{pmatrix}1&0\\0&e^{i\alpha_3}\end{pmatrix}\begin{pmatrix}\cos{\frac{\alpha_2}{2}}&-i\sin{\frac{\alpha_2}{2}}\\-i\sin{\frac{\alpha_2}{2}}&\cos{\frac{\alpha_2}{2}}\end{pmatrix}\begin{pmatrix}1&0\\0&e^{i\alpha_1}\end{pmatrix}\\
&= e^{i(\gamma+\frac{\alpha_2}{2})}\begin{pmatrix}1&0\\0&e^{i(\alpha_3+\frac{\pi}{2})}\end{pmatrix}\begin{pmatrix}\cos{\frac{\alpha_2}{2}}&\sin{\frac{\alpha_2}{2}}\\-\sin{\frac{\alpha_2}{2}}&\cos{\frac{\alpha_2}{2}}\end{pmatrix}\begin{pmatrix}1&0\\0&e^{i(\alpha_1-\frac{\pi}{2})}\end{pmatrix}\\
&= e^{i(\gamma+\frac{\alpha_1+\alpha_2+\alpha_3}{2})}\begin{pmatrix}e^{-i(\frac{\alpha_3}{2}+\frac{\pi}{4})}&0\\0&e^{i(\frac{\alpha_3}{2}+\frac{\pi}{4})}\end{pmatrix}\begin{pmatrix}\cos{\frac{\alpha_2}{2}}&\sin{\frac{\alpha_2}{2}}\\-\sin{\frac{\alpha_2}{2}}&\cos{\frac{\alpha_2}{2}}\end{pmatrix}\begin{pmatrix}e^{-i(\frac{\alpha_1}{2}-\frac{\pi}{4})}&0\\0&e^{i(\frac{\alpha_1}{2}-\frac{\pi}{4})}\end{pmatrix}
\end{align*}
$\bullet$ Uniqueness:\\
Suppose $\interp{
\input{./figures/one-qubit-unitary.tikz}
}=\interp{
\input{./figures/one-qubit-unitary-2.tikz}
}$. The first diagram yields:
\[e^{i(\gamma+\frac{\alpha_2}{2})}\begin{pmatrix}\cos{\frac{\alpha_2}{2}}&-ie^{i\alpha_1}\sin{\frac{\alpha_2}{2}}\\-ie^{i\alpha_3}\sin{\frac{\alpha_2}{2}}&e^{i(\alpha_1+\alpha_3)}\cos{\frac{\alpha_2}{2}}\end{pmatrix}\]
and similarly for the second one. If $\alpha_2\neq 0\bmod \pi$, then neither $\cos{\frac{\alpha_2}{2}}$ nor $\sin{\frac{\alpha_2}{2}}$ is null. Hence, dividing element (1,1) by element (0,0) on both sides gives $e^{i(\alpha_1+\alpha_3)}=e^{i(\alpha_1'+\alpha_3')}$ and dividing element (0,1) by element (1,0) on both sides gives $e^{i(\alpha_1-\alpha_3)}=e^{i(\alpha_1'-\alpha_3')}$. In other words, $\alpha_1+\alpha_3=\alpha_1'+\alpha_3'\bmod 2\pi$ and $\alpha_1-\alpha_3=\alpha_1'-\alpha_3'\bmod 2\pi$, so $2\alpha_1=2\alpha_1'\bmod 2\pi$ i.e.~$\alpha_1=\alpha_1'\bmod \pi$. Since we required $\alpha_1,\alpha_1'\in[0,\pi)$, we get $\alpha_1=\alpha_1'$. It then follows easily that $\alpha_3=\alpha_3'$, $\alpha_2=\alpha_2'$ and $\gamma=\gamma'$.
\end{proof}

\begin{proof}[Necessity of Rule \eu]
\phantomsection \label{prf:hd-necessity}
We define the non-standard interpretation $\interp{.}^{\natural}$ as follows:
$$
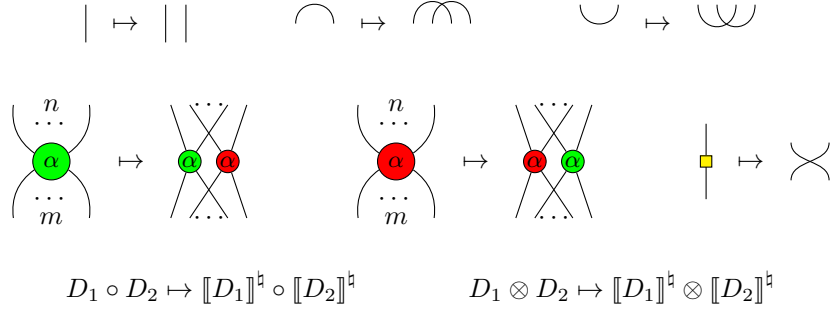
\input{./figures/single-line.tikz}
~~\mapsto ~~
\input{./figures/single-line.tikz}
~~
\input{./figures/single-line.tikz}
\qquad\qquad
\input{./figures/caps.tikz}
~~\mapsto~~
\input{./figures/2-caps.tikz}
\qquad\qquad
\input{./figures/cup.tikz}
~~\mapsto~~
\input{./figures/2-cups.tikz}
$$
$$
\input{./figures/gn-alpha.tikz}
~~\mapsto~~
\input{./figures/hd-proof-gn-alpha.tikz}
\qquad\qquad
\input{./figures/rn-alpha.tikz}
~~\mapsto~~
\input{./figures/hd-proof-rn-alpha.tikz}
\qquad\qquad
\input{./figures/Hadamard.tikz}
~~\mapsto~~
\input{./figures/crossing.tikz}
$$
$$D_1\circ D_2\mapsto \interp{D_1}^{\natural}\circ\interp{D_2}^{\natural} \qquad\qquad D_1\otimes D_2\mapsto \interp{D_1}^{\natural}\otimes\interp{D_2}^{\natural} $$
It is then easy to see that all the rules but \eu hold under this interpretation, hence proving that \eu could not be derived from the other rules.
\end{proof}

\subsection{Proofs for Clifford}

\begin{proof}[Lemma \ref{lem:hopf}]
\def\fig{hopf}
\begin{align*}
\input{./figures/\fig/\fig_00.tikz}
\eq{}\input{./figures/\fig/\fig_01.tikz}
\eq{\stt\\\soo}\input{./figures/\fig/\fig_02.tikz}
\eq{\bt}\input{./figures/\fig/\fig_03.tikz}
\eq{\bo}\input{./figures/\fig/\fig_04.tikz}
\eq{\soo\\\stt}\input{./figures/\fig/\fig_05.tikz}
\end{align*}
\end{proof}

\begin{proof}[Prop. \ref{prop:inverse}]
\def\fig{scalar-inverse-proof}
\begin{align*}
\input{./figures/\fig/\fig_00.tikz}
\eq{\e}\input{./figures/\fig/\fig_01.tikz}
\eq{\soo\\\bo}\input{./figures/\fig/\fig_02.tikz}
\eq{\soo}\input{./figures/\fig/\fig_03.tikz}
\eq{\ref{lem:hopf}}\input{./figures/\fig/\fig_04.tikz}\\
\eq{\soo\\\stt}\input{./figures/\fig/\fig_05.tikz}
\eq{\soo\\\stt}\input{./figures/\fig/\fig_06.tikz}
\eq{\e}\input{./figures/\fig/\fig_07.tikz}
\end{align*}
\end{proof}

\begin{proof}[Prop. \ref{prop:pi-commutation}]
\def\fig{k2s-proof}
\begin{align*}
\input{./figures/\fig/\fig_00.tikz}
\eq{\stt\\\soo}\input{./figures/\fig/\fig_01.tikz}
\eq{\euler}\input{./figures/\fig/\fig_02.tikz}
\eq{\stt\\\soo\\\ref{prop:inverse}}\input{./figures/\fig/\fig_03.tikz}
\end{align*}
\end{proof}

\begin{proof}[Prop. \ref{prop:zero}]
\def\fig{2-is-sqrt-2-squared}
\begin{align}
\input{./figures/\fig/\fig_00.tikz}
\eq{\soo\\\stt}\input{./figures/\fig/\fig_01.tikz}
\eq{\bo}\input{./figures/\fig/\fig_02.tikz}
\eq{\soo}\input{./figures/\fig/\fig_03.tikz}
\label{eq:2-is-sqrt-2-squared}
\end{align}
\def\fig{scalar-bicolor-0-pi}
\begin{align}
\input{./figures/\fig/\fig_00.tikz}
\eq{\soo}\input{./figures/\fig/\fig_01.tikz}
\eq{\ref{prop:pi-commutation}}\input{./figures/\fig/\fig_02.tikz}
\eq{\stt}\input{./figures/\fig/\fig_03.tikz}
\eq{\bo\\\ref{prop:inverse}}\input{./figures/\fig/\fig_04.tikz}
\eq{\soo}\input{./figures/\fig/\fig_05.tikz}
\label{eq:scalar-bicolor-0-pi}
\end{align}
\def\fig{zero-proof-1}
\begin{align}
\input{./figures/\fig/\fig_00.tikz}
\eq{\ref{lem:hopf}}\input{./figures/\fig/\fig_01.tikz}
\eq{\soo}\input{./figures/\fig/\fig_02.tikz}
\eq{\ref{prop:pi-commutation}}\input{./figures/\fig/\fig_03.tikz}
\eq{\soo}\input{./figures/\fig/\fig_04.tikz}
\eq{\ref{prop:inverse}\\\ref{lem:hopf}}\input{./figures/\fig/\fig_05.tikz}
\label{eq:zero-1}
\end{align}
\def\fig{zero-proof-2}
\begin{align}
\input{./figures/\fig/\fig_00.tikz}
\eq{(\ref{eq:zero-1})}\input{./figures/\fig/\fig_01.tikz}
\eq{(\ref{eq:scalar-bicolor-0-pi})}\input{./figures/\fig/\fig_02.tikz}
\eq{\ref{prop:inverse}}\input{./figures/\fig/\fig_03.tikz}
\label{eq:zero-2}
\end{align}
Now, if $\alpha\in\mathbb{D}\pi$ (where $\mathbb{D}:=\mathbb{Z}\left[\frac{1}{2}\right]$), then there exists $n$ such that $2^n\alpha=0\bmod 2\pi$. Hence, in this case the scalar on the right hand side of (\ref{eq:zero-1}) can be removed by applying (\ref{eq:zero-2}) from right to left $n+1$ times then using (\ref{eq:scalar-bicolor-0-pi}) and \ref{prop:inverse} to remove it. Hence:
\begin{equation}
\forall \alpha \in \mathbb{D}\pi,~~
\input{./figures/zero-proof-aux.tikz}

\label{eq:zero-angle}
\end{equation}
\def\fig{zero-proof-3}
\begin{align}
\input{./figures/\fig/\fig_00.tikz}
\eq{(\ref{eq:zero-angle})}\input{./figures/\fig/\fig_01.tikz}
\eq{\e}\input{./figures/\fig/\fig_02.tikz}
\label{eq:zero-sqrt-2}
\end{align}
\def\fig{rn-0-1-pi_2-to-gn}
\begin{align}
\input{./figures/\fig/\fig_00.tikz}
\eq{\h}\input{./figures/\fig/\fig_01.tikz}
\eq{\eu\\\soo\\\ref{prop:inverse}}\input{./figures/\fig/\fig_02.tikz}
\eq{\bo\\\soo}\input{./figures/\fig/\fig_03.tikz}
\label{eq:rn-0-1-pi_2-to-gn}
\end{align}
\def\fig{zero-proof-4}
\begin{align*}
\input{./figures/\fig/\fig_00.tikz}
\eq{(\ref{eq:zero-angle})}\input{./figures/\fig/\fig_01.tikz}
\eq{(\ref{eq:rn-0-1-pi_2-to-gn})}\input{./figures/\fig/\fig_02.tikz}
\eq{(\ref{eq:zero-angle})}\input{./figures/\fig/\fig_03.tikz}
\eq{(\ref{eq:2-is-sqrt-2-squared})\\\ref{prop:inverse}}\input{./figures/\fig/\fig_04.tikz}
\eq{(\ref{eq:zero-sqrt-2})}\input{./figures/\fig/\fig_05.tikz}
\end{align*}
\end{proof}

\begin{proof}[Lem. \ref{lem:multiplying-phases}]
We are going to use the following equation we get from the Clifford completeness:
\def\fig{pi-distribution}
\begin{align}
\label{eq:pi-distrib}
\input{./figures/\fig/\fig_00.tikz}
\eq{}\input{./figures/\fig/\fig_06.tikz}
\end{align}
so:
\def\fig{multiplying-global-phases-proof}
\begin{align*}
\input{./figures/\fig/\fig_00.tikz}
\eq{\soo\\\bo}\input{./figures/\fig/\fig_01.tikz}
\eq{\ref{eq:pi-distrib}\\\soo}\input{./figures/\fig/\fig_02.tikz}
\eq{\soo}\input{./figures/\fig/\fig_03.tikz}
\end{align*}
\end{proof}

\begin{proof}[Lem. \ref{lem:bicolor-0-alpha}]
\def\fig{bicolor-0-alpha-bis}
\begin{align*}
\input{./figures/\fig/\fig_00.tikz}
\eq{\soo}\input{./figures/\fig/\fig_01.tikz}
\eq{\ref{prop:pi-commutation}}\input{./figures/\fig/\fig_02.tikz}
\eq{\soo\\\bo}\input{./figures/\fig/\fig_03.tikz}
\eq{\ref{lem:multiplying-phases}\\\ref{prop:inverse}}\input{./figures/\fig/\fig_04.tikz}
\eq{\bo}\input{./figures/\fig/\fig_05.tikz}
\eq{\soo\\\ref{prop:inverse}}\input{./figures/\fig/\fig_06.tikz}
\end{align*}
\end{proof}

\subsection{Proofs for Singular-Value Decompositions}

\begin{proof}[Lem. \ref{lem:euler-branch}]
\def\fig{Euler-branch}
\begin{align*}
\input{./figures/\fig/\fig_00.tikz}
\eq{\bt}\input{./figures/\fig/\fig_01.tikz}
\eq{\soo\\\euler}\input{./figures/\fig/\fig_02.tikz}
\eq{\bt}\input{./figures/\fig/\fig_03.tikz}
\end{align*}
\end{proof}

\begin{proof}[Cor. \ref{cor:euler-branch}]
\def\fig{Euler-branch-cor-2}
\begin{align*}
\input{./figures/\fig/\fig_00.tikz}
\eq{\soo}\input{./figures/\fig/\fig_01.tikz}
\eq{\comp2}\input{./figures/\fig/\fig_02.tikz}
\eq{\ref{lem:euler-branch}}\input{./figures/\fig/\fig_03.tikz}
\end{align*}
\end{proof}

\begin{proof}[Lem. \ref{lem:two-branches-to-one}]
\def\fig{two-branches-to-one-by-euler-2}
\begin{align*}
\input{./figures/\fig/\fig_00.tikz}
\eq{\bt}\input{./figures/\fig/\fig_01.tikz}
\eq{\comp2}\input{./figures/\fig/\fig_02.tikz}
\eq{\euler\\\soo}\input{./figures/\fig/\fig_03.tikz}
\eq{\h\\\soo}\input{./figures/\fig/\fig_04.tikz}\\
\eq{\comp2}\input{./figures/\fig/\fig_05.tikz}
\eq{\bt}\input{./figures/\fig/\fig_06.tikz}
\end{align*}
\end{proof}

\begin{proof}[Lem. \ref{lem:sup-gen}]
\def\fig{Euler-branch-to-line-3}
\begin{align*}
\input{./figures/\fig/\fig_00.tikz}
\eq{\soo}\input{./figures/\fig/\fig_01.tikz}
\eq{\ref{cor:euler-branch}}\input{./figures/\fig/\fig_02.tikz}
\eq{\bo\\\soo\\\stt}\input{./figures/\fig/\fig_03.tikz}
\end{align*}
\end{proof}

\begin{proof}[Prop. \ref{prop:supp}]
We first use Lemma \ref{lem:sup-gen}, where $\alpha_3=\alpha_1+\pi$. In this case, it can be computed that $\beta_1+\beta_3=0$, so we end up with:\def\fig{former-supp-proof-2}$\input{./figures/\fig/\fig_00.tikz}\eq{}\input{./figures/\fig/\fig_01.tikz}$. From this, we can easily specify the scalar on the right part:
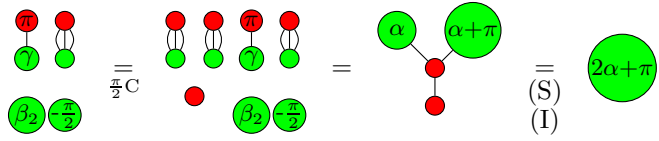
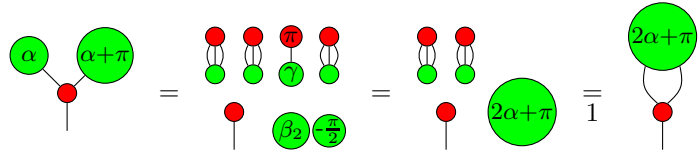
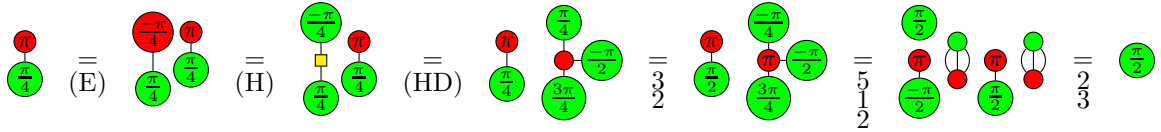
\begin{align*}
\input{./figures/\fig/\fig_02.tikz}
\eq{\comp2}\input{./figures/\fig/\fig_03.tikz}
\eq{}\input{./figures/\fig/\fig_04.tikz}
\eq{\soo\\\stt}\input{./figures/\fig/\fig_05.tikz}
\end{align*}
So finally:
\begin{align*}
\input{./figures/\fig/\fig_06.tikz}
\eq{}\input{./figures/\fig/\fig_07.tikz}
\eq{}\input{./figures/\fig/\fig_08.tikz}
\eq{\ref{lem:hopf}}\input{./figures/\fig/\fig_09.tikz}
\end{align*}
\end{proof}


\begin{proof}[Rem.~\ref{rem:gn-pi_2-0-0-equals-sqrt2-exp-pi_4}]
\def\fig{gn-pi_2-0-0-equals-sqrt2-exp-pi_4-proof}
\begin{align*}
\input{./figures/\fig/\fig_00.tikz}
\eq{\e}\input{./figures/\fig/\fig_01.tikz}
\eq{\h}\input{./figures/\fig/\fig_02.tikz}
\eq{\eu}\input{./figures/\fig/\fig_03.tikz}
\eq{\ref{prop:pi-commutation}\\\ref{lem:multiplying-phases}}\input{./figures/\fig/\fig_04.tikz}
\eq{\ref{prop:supp}\\\ref{lem:hopf}\\\ref{prop:inverse}}\input{./figures/\fig/\fig_05.tikz}
\eq{\ref{lem:multiplying-phases}\\\ref{lem:bicolor-0-alpha}}\input{./figures/\fig/\fig_06.tikz}
\end{align*}
\end{proof}

\begin{proof}[Prop. \ref{prop:state-SVD}]
First, notice that a state in the previous form can easily be transformed into an SVD. Indeed, if $\beta\in[\pi,2\pi)$:
\def\fig{state-to-SVD}
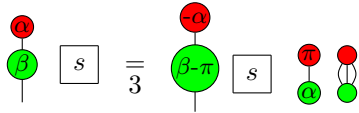
\begin{align*}
\input{./figures/\fig/\fig_00.tikz}\eq{\ref{prop:pi-commutation}}\input{./figures/\fig/\fig_01.tikz}
\end{align*}
and similarly for the SVD$_r$. We can show that we can transform an SVD$_r$ into an SVD$_g$ and vice-versa:
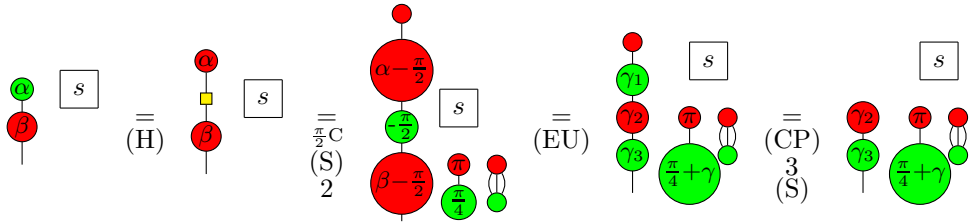
\def\fig{state-SVD-g-to-r}:
\begin{align*}
\input{./figures/\fig/\fig_00.tikz}
\eq{\h}\input{./figures/\fig/\fig_01.tikz}
\eq{\comp2\\\soo\\\ref{rem:gn-pi_2-0-0-equals-sqrt2-exp-pi_4}}\input{./figures/\fig/\fig_02.tikz}
\eq{\euler}\input{./figures/\fig/\fig_03.tikz}
\eq{\bo\\\ref{lem:bicolor-0-alpha}\\\soo}\input{./figures/\fig/\fig_04.tikz}
\end{align*}
Then, we prove the result by induction.
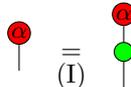
\def\fig{rn-0-1-to-generic}
\[\input{./figures/\fig/\fig_00.tikz}\eq{\stt}\input{./figures/\fig/\fig_01.tikz}\]
Then \def\fig{h-to-generic}:
\begin{align*}
\input{./figures/\fig/\fig_00.tikz}
\eq{}\input{./figures/\fig/\fig_01.tikz}
\eq{\h}\input{./figures/\fig/\fig_02.tikz}
\eq{}\input{./figures/\fig/\fig_03.tikz}
\end{align*}
Notice that the generator $R_Z^{(0,1)}(\alpha)$ can be obtained as a combination of the last two.
Then \def\fig{g-1-1-to-generic}:
\begin{align*}
\input{./figures/\fig/\fig_00.tikz}
\eq{}\input{./figures/\fig/\fig_01.tikz}
\eq{\soo}\input{./figures/\fig/\fig_02.tikz}
\end{align*}
\def\fig{g-2-1-to-generic}
\begin{align*}
\input{./figures/\fig/\fig_00.tikz}
&\eq{\soo}\input{./figures/\fig/\fig_01.tikz}
\eq{\ref{lem:sup-gen}}\input{./figures/\fig/\fig_02.tikz}\\
&\eq{\comp2}\input{./figures/\fig/\fig_03.tikz}
\eq{\h}\input{./figures/\fig/\fig_04.tikz}
\end{align*}
Finally, the generator $R_Z^{(n,1)}(\alpha)$ can be obtained by composition of $R_Z(\alpha)$ and $R_Z^{(2,1)}(\alpha)$; and $R_X^{(n,1)}(\alpha)$ can be obtained by composition of $R_Z^{(n,1)}(\alpha)$ and $H$.
\end{proof}

\begin{proof}[Prop. \ref{prop:state-SVD-uniqueness}]
The equality reads $s_1\begin{pmatrix}1+e^{i\alpha_1}\\e^{i\beta_1}(1-e^{i\alpha_1})\end{pmatrix}=s_2\begin{pmatrix}1+e^{i\alpha_2}\\e^{i\beta_2}(1-e^{i\alpha_2})\end{pmatrix}$. If $\alpha_1=\pi$, then it is easy to see that $\alpha_2=\pi$ and $s_1e^{i\beta_1}=s_2e^{i\beta_2}$. If $\alpha_i\neq\pi$, then the upper coefficient is non-null, hence we can divide the lower coefficient by the upper one, which yields:
\begin{align*}
e^{i\beta_1}\frac{1-e^{i\alpha_1}}{1+e^{i\alpha_1}} = e^{i\beta_2}\frac{1-e^{i\alpha_2}}{1+e^{i\alpha_2}} 
& \qquad\Leftrightarrow\qquad e^{i\beta_1}\tan{\frac{\alpha_1}{2}}=e^{i\beta_2}\tan{\frac{\alpha_2}{2}}
\end{align*}
If $\alpha_1=0\bmod 2\pi$ then $\alpha_2 = 0\bmod 2\pi$. Otherwise, 
since $\beta_1,\beta_2\in[0,\pi)$, $\beta_1=\beta_2$ and $\alpha_1=\alpha_2\bmod 2\pi$.
\end{proof}

\begin{proof}[Prop. \ref{prop:1-1-diag-SVD}]
First, if $D$ is in the form SVD$_g$, but where the constraints on the angles are not met, we can transform it into an actual SVD$_g$:
\begin{itemize}
\item If $\alpha_1\in[\pi,2\pi)$ (and similarly for $\alpha_5$):
\def\fig{1-1-SVD-alpha-pi-2pi}
\begin{align*}
\input{./figures/\fig/\fig_00.tikz}
\eq{\soo\\\ref{prop:pi-commutation}}\input{./figures/\fig/\fig_01.tikz}
\end{align*}
\item If $\gamma \in [-\frac{\pi}{2},0[$:
\def\fig{1-1-SVD-gamma-minus-pi_2-0}
\begin{align*}
\input{./figures/\fig/\fig_00.tikz}
\eq{\soo\\\ref{prop:pi-commutation}}\input{./figures/\fig/\fig_01.tikz}
\end{align*}
\item If $\gamma \in [-\pi,-\frac{\pi}{2}[$:
\def\fig{1-1-SVD-gamma-minus-pi-minus-pi_2}
\begin{align*}
\input{./figures/\fig/\fig_00.tikz}
\eq{\soo\\\ref{prop:pi-commutation}}\input{./figures/\fig/\fig_01.tikz}
\end{align*}
\item If $\gamma \in [\frac{\pi}{2},\pi)$:
\def\fig{1-1-SVD-gamma-pi_2-pi}
\begin{align*}
\input{./figures/\fig/\fig_00.tikz}
\eq{\soo\\\ref{prop:pi-commutation}}\input{./figures/\fig/\fig_01.tikz}
\eq{\soo\\\ref{prop:pi-commutation}}\input{./figures/\fig/\fig_02.tikz}
\end{align*}
\end{itemize}

Then, we show that the two decompositions are equivalent:
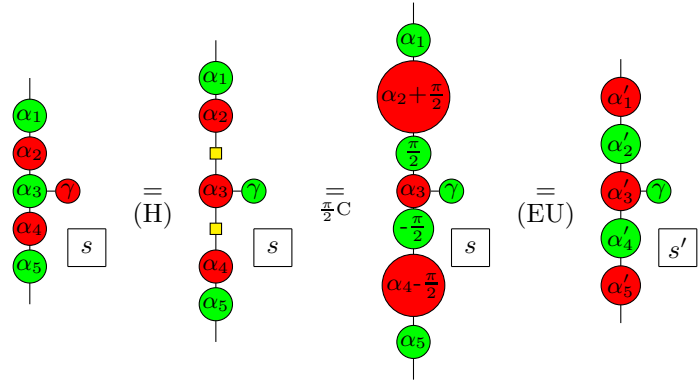
\def\fig{1-1-diagram-decomposition-by-singular-values}
\begin{align*}
\input{./figures/\fig/\fig_01.tikz}
\eq{\h}\input{./figures/\fig/\fig_02.tikz}
\eq{\comp2}\input{./figures/\fig/\fig_03.tikz}
\eq{\euler}\input{./figures/\fig/\fig_04.tikz}
\end{align*}
The two $1\to 1$ generators $R_Z^{(1,1)}(\alpha)$ and $H$ can be put in SVD:
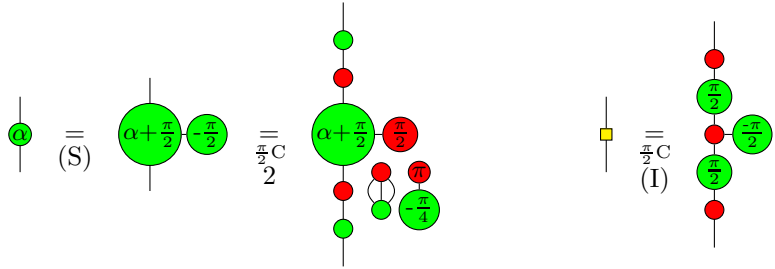
\def\fig{gn-1-1-SVD}
\begin{align*}
\input{./figures/\fig/\fig_00.tikz}
\eq{\soo}\input{./figures/\fig/\fig_01.tikz}
\eq{\comp2\\\ref{rem:gn-pi_2-0-0-equals-sqrt2-exp-pi_4}}\input{./figures/\fig/\fig_02.tikz}
\qquad\qquad\qquad\def\fig{hadamard-SVD}
\input{./figures/\fig/\fig_00.tikz}
\eq{\comp2\\\stt}\input{./figures/\fig/\fig_01.tikz}
\end{align*}
The composition of two SVDs can be put in SVD (here, ignoring the scalars):
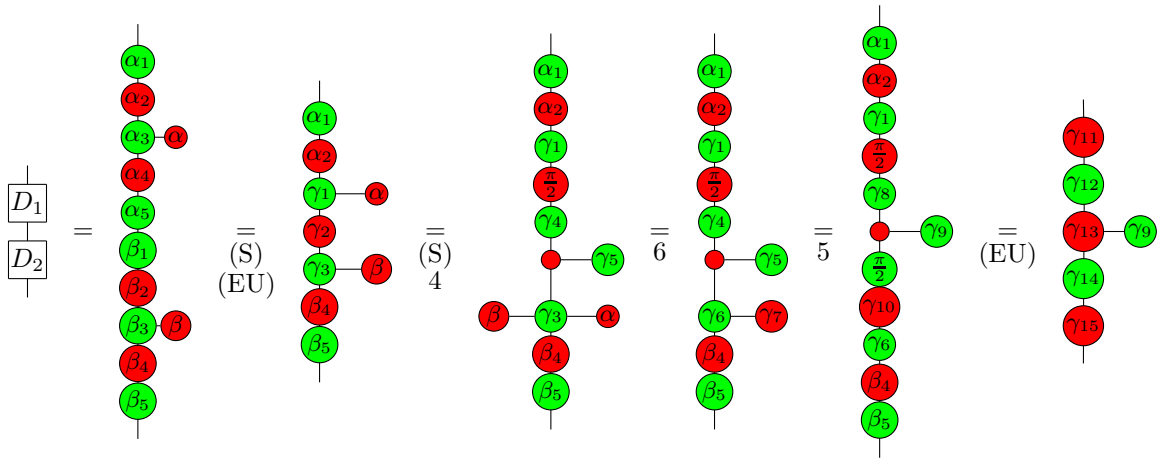
\def\fig{1-1-diagram-composition-SVD}
\begin{align*}
\input{./figures/\fig/\fig_00.tikz}
\eq{}\input{./figures/\fig/\fig_01.tikz}
\eq{\soo\\\euler}\input{./figures/\fig/\fig_02.tikz}
\eq{\soo\\\ref{lem:euler-branch}}\input{./figures/\fig/\fig_03.tikz}
\eq{\ref{lem:sup-gen}}\input{./figures/\fig/\fig_04.tikz}
\eq{\ref{lem:two-branches-to-one}}\input{./figures/\fig/\fig_05.tikz}
\eq{\euler}\input{./figures/\fig/\fig_06.tikz}
\end{align*}
Notice that, by composition, the $1\to 1$ generator $R_X^{(1,1)}(\alpha)$ can be put in SVD.

If the $1\to 1$ diagram has no cycle, there can still be branching. Hence, there can be a state $D:0\to 1$ in tree-like form attached to the ``main wire'' by a node, say green, as follows:
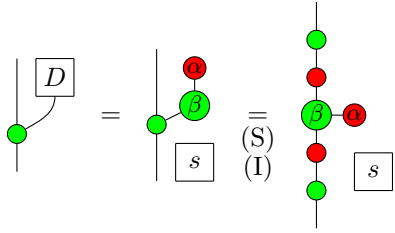
\def\fig{1-1-diagram-with-branching-SVD}
\begin{align*}
\input{./figures/\fig/\fig_00.tikz}
\eq{}\input{./figures/\fig/\fig_01.tikz}
\eq{\soo\\\stt}\input{./figures/\fig/\fig_02.tikz}
\end{align*}
Branching made by a red node can be deduced by composing the green one and Hadamard nodes.
\end{proof}

\begin{proof}[Prop. \ref{prop:1-1-SVD-essentially-unique}]
First we decompose $D_1$ and $D_2$ as:
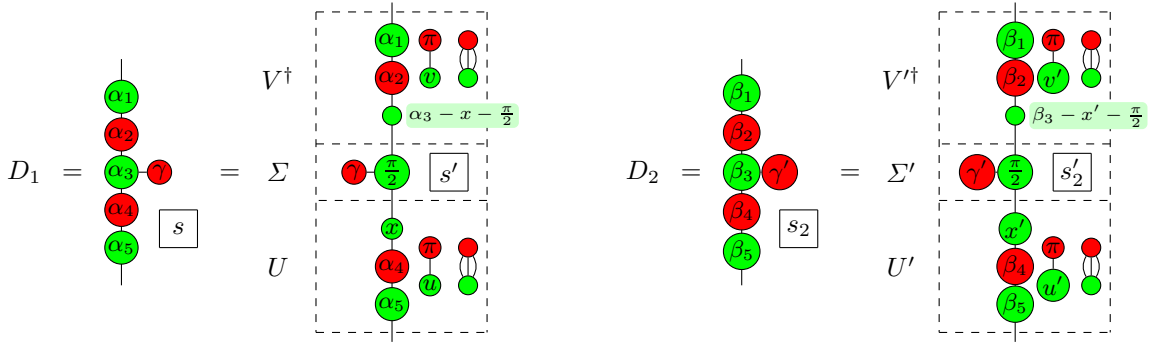
\[\def\fig{1-1-diagram-decomposition-by-singular-values}D_1\eq{}\input{./figures/\fig/\fig_01.tikz}\eq{}
\input{./figures/1-1-SVD-uniqueness-D1-decomp.tikz}
\qquad\qquad D_2\eq{}
\input{./figures/D2-SVD.tikz}
\eq{}
\input{./figures/1-1-SVD-uniqueness-D2-decomp.tikz}
\]
where $u$, $v$, $u'$ and $v'$ have been chosen so that $\interp{\Sigma}$ and $\interp{\Sigma'}$ are real matrices. Notice that $\interp{U}$, $\interp{V^{\dagger}}$, $\interp{U'}$, $\interp{V'^{\dagger}}$ are unitaries. We have two SVDs that represent the same matrix:
\[\interp{U}\circ\interp{\Sigma}\circ\interp{V^{\dagger}} = \interp{D_1} = \interp{D_2} = \interp{U'}\circ\interp{\Sigma'}\circ\interp{V'^{\dagger}}\]
First off, let us show that $\Sigma$ and $\Sigma'$ are essentially the same. One could compute $\interp{\Sigma} = \interp{s'}(1+e^{i\gamma})\begin{pmatrix}1&0\\0&\tan{\frac{\gamma}{2}}\end{pmatrix}$ and $\interp{\Sigma'} = \interp{s_2'}(1+e^{i\gamma'})\begin{pmatrix}1&0\\0&\tan{\frac{\gamma'}{2}}\end{pmatrix}$. Since $\gamma,\gamma'\in[0,\frac{\pi}{2}]$, $\tan{\frac{\gamma}{2}}$ and $\tan{\frac{\gamma'}{2}}$ are smaller than $1$, and since the diagrams are non-null, we get $\interp{\Sigma}=\interp{\Sigma'}$ by Proposition \ref{prop:SVD-properties}, which implies $\gamma=\gamma'$.

If $\gamma = \gamma' \neq 0$, then $\interp{\Sigma}$ and $\interp{\Sigma'}$ have full rank. Moreover, if $\gamma = \gamma' \neq \frac{\pi}{2}$, then $\interp{\Sigma}$ and $\interp{\Sigma'}$ are not colinear to the identity. Hence, if $\gamma = \gamma' \in ]0,\frac{\pi}{2}[$, then we can apply Proposition \ref{prop:SVD-properties}.

By Proposition \ref{prop:SVD-properties}, there exists $d=\begin{pmatrix}e^{i\varphi_0}&0\\0&e^{i\varphi_1}\end{pmatrix}$ such that $\interp{U'}=\interp{U}\circ d$ and $\interp{V'^{\dagger}} = d^{\dagger}\circ\interp{V^{\dagger}}$. Notice that $\interp{
\input{./figures/diagram-d-for-SVD.tikz}
}=d$ and $\interp{
\input{./figures/diagram-d-dagger-for-SVD.tikz}
}=d^{\dagger}$. Hence:
\def\fig{Up-and-U-d-for-SVD}
\begin{align*}
\interp{U'}=\interp{\input{./figures/\fig/\fig_00.tikz}}=\interp{U}\circ d = \interp{\input{./figures/\fig/\fig_01.tikz}}
\end{align*}
Since $\beta_5$ and $\alpha_5$ are in $[0,\pi)$, the representation of the unitary is unique by Proposition \ref{prop:unitary-uniqueness}, so $\beta_5=\alpha_5$, $\beta_4=\alpha_4$, and $x'=x+\varphi_1-\varphi_0$. Similarly, the second equation yields $\alpha_1=\beta_1$, $\alpha_2 = \beta_2$ and $\beta_3-x'-\frac{\pi}{2} = \alpha_3-x-\frac{\pi}{2}+\varphi_0-\varphi_1$. Together, the equations on $x$ and $x'$ imply that $\alpha_3=\beta_3$.
\end{proof}

\subsection{Proofs for Clifford+T and Beyond}

\begin{proof}[Prop. \ref{prop:BW}]
Using Proposition \ref{prop:1-1-diag-SVD}, we can put both sides of the equation in SVD, and thanks to Proposition \ref{prop:1-1-SVD-essentially-unique}, the two forms have the same angles. We can even compute:
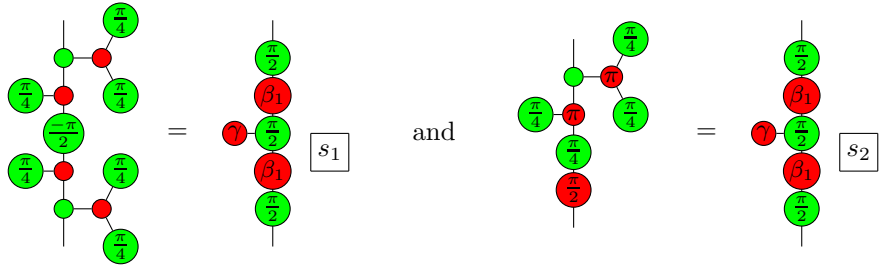
\[
\input{./figures/BW-LHS-SVD.tikz}
\qquad\text{ and }\qquad

\input{./figures/BW-RHS-SVD.tikz}
\]
with $\gamma = \frac{\pi}{2}-2\arctan{\frac{1}{\sqrt{5}}}$ and $\beta_1 = \arctan{2}$.
It remains to prove that the two scalar diagrams are equal:
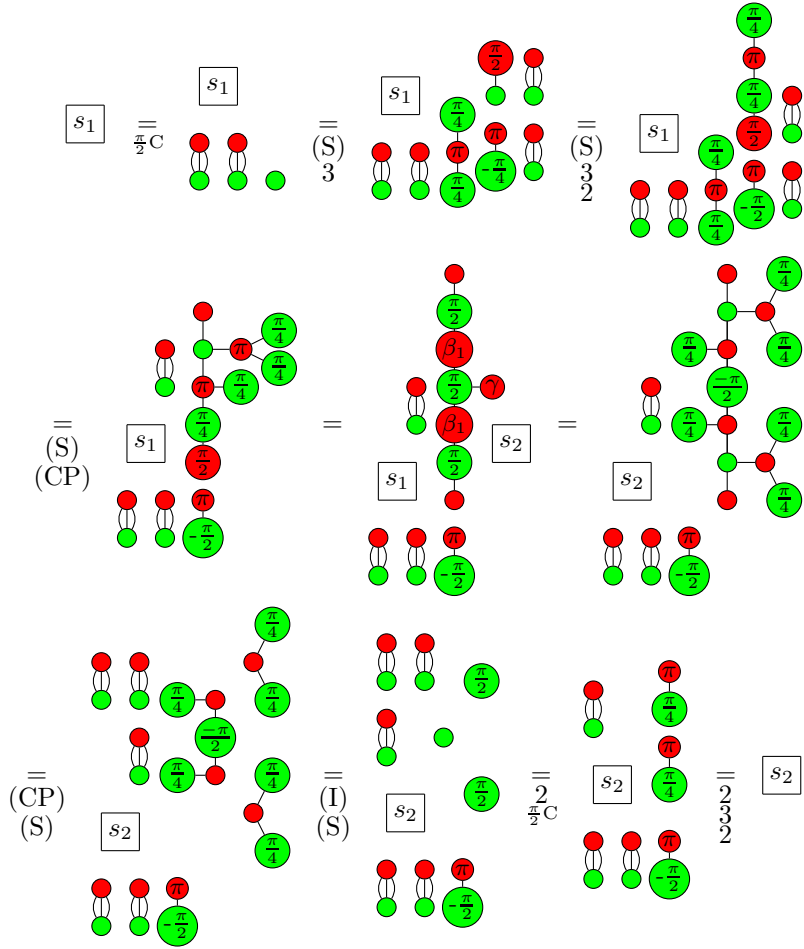
\def\fig{BW-proof-scalars}
\begin{align*}
\input{./figures/\fig/\fig_00.tikz}
\eq{\comp2}\input{./figures/\fig/\fig_01.tikz}
\eq{\soo\\\ref{prop:pi-commutation}}\input{./figures/\fig/\fig_02.tikz}
\eq{\soo\\\ref{prop:pi-commutation}\\\ref{lem:multiplying-phases}}\input{./figures/\fig/\fig_03.tikz}\\
\eq{\soo\\\bo}\input{./figures/\fig/\fig_04.tikz}
\eq{}\input{./figures/\fig/\fig_05.tikz}
\eq{}\input{./figures/\fig/\fig_06.tikz}\\
\eq{\bo\\\soo}\input{./figures/\fig/\fig_07.tikz}
\eq{\stt\\\soo}\input{./figures/\fig/\fig_08.tikz}
\eq{\ref{rem:gn-pi_2-0-0-equals-sqrt2-exp-pi_4}\\\comp2}\input{./figures/\fig/\fig_09.tikz}
\eq{\ref{lem:multiplying-phases}\\\ref{lem:bicolor-0-alpha}\\\ref{prop:inverse}}\input{./figures/\fig/\fig_10.tikz}
\end{align*}
\end{proof}

\begin{proof}[Lem. \ref{lem:C1-h}]
We prove the equality by simplifying both sides of the equation. The left hand side yields, when ignoring the scalars:
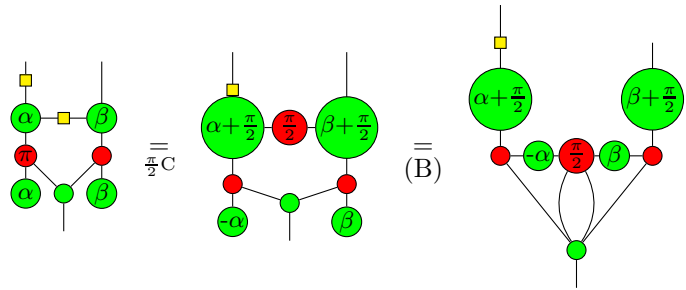
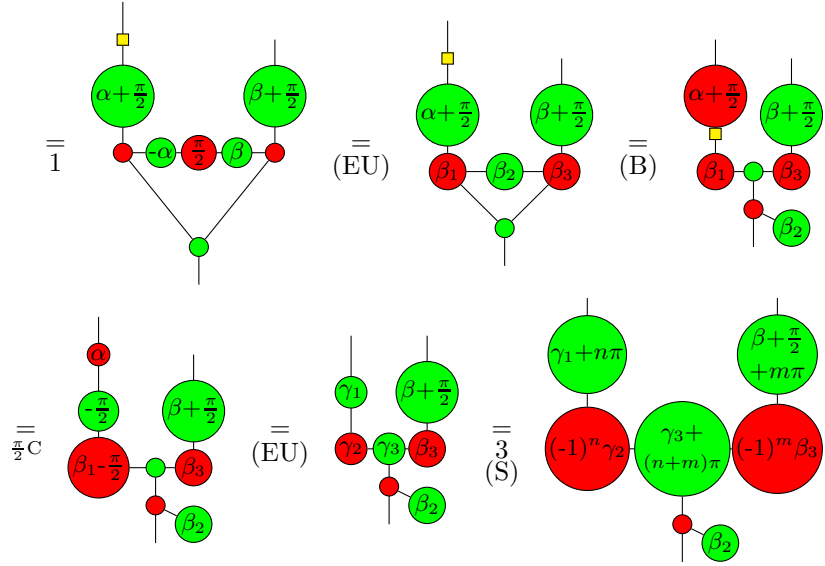
\def\fig{C1-proof}
\begin{align*}
\input{./figures/\fig/\fig_00.tikz}
\eq{\comp2}\input{./figures/\fig/\fig_01.tikz}
\eq{\bt}\input{./figures/\fig/\fig_02.tikz}\\
\eq{\ref{lem:hopf}}\input{./figures/\fig/\fig_03.tikz}
\eq{\euler}\input{./figures/\fig/\fig_04.tikz}
\eq{\bt}\input{./figures/\fig/\fig_05.tikz}\\
\eq{\comp2}\input{./figures/\fig/\fig_06.tikz}
\eq{\euler}\input{./figures/\fig/\fig_07.tikz}
\eq{\ref{prop:pi-commutation}\\\soo}\input{./figures/\fig/\fig_08.tikz}
\end{align*}
where $n$ and $m$ are chosen in $\{0,1\}$ so that $\gamma_1+n\pi$ and $\beta+\frac{\pi}{2}+m\pi$ are in $[0,\pi)$.
Similarly, the right hand side yields:
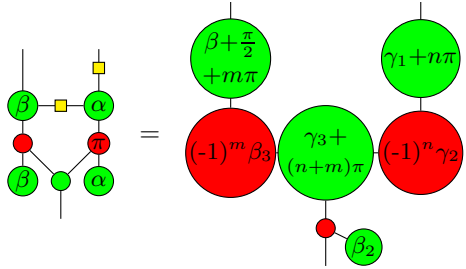
\def\fig{C1-proof-2}
\begin{align*}
\input{./figures/\fig/\fig_00.tikz}
\eq{}\input{./figures/\fig/\fig_01.tikz}
\end{align*}
Notice that, due to the symmetry of the two diagrams, the resulting scalars (that we ignored) are equal (and non null). If $\beta_2=0\bmod \pi$, then we can compute that both $\alpha$ and $\beta$ are multiples of $\pi$, and in this case the equation is trivially derivable. Else,
notice that $\interp{
\input{./figures/C1-proof-branch-to-remove.tikz}
}$ is invertible, $\left(\text{its inverse is } \frac{1}{1-e^{2i\beta_2}}\begin{pmatrix}1 & -e^{i\beta_2}\\-e^{i\beta_2}&1\end{pmatrix}\right)$. Hence, we get:
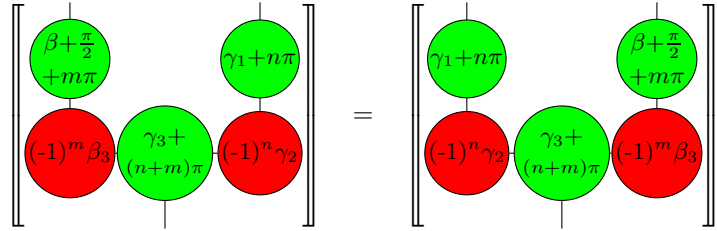
\def\fig{C1-proof-aux}
\begin{align*}
\interp{\input{./figures/\fig/\fig_00.tikz}}
\eq{}\interp{\input{./figures/\fig/\fig_01.tikz}}
\end{align*}
We can then plug any red dot with angle $\in]0,\frac{\pi}{2}[$, say $\frac{\pi}{4}$, on the lower branch. We can now use Proposition \ref{prop:1-1-SVD-essentially-unique}, match the angles $\gamma_1+n\pi = \beta+\frac{\pi}{2}+m\pi$ and $(-1)^n\gamma_2 = (-1)^m\beta_3$, so the two initial diagrams are equal.
\end{proof}

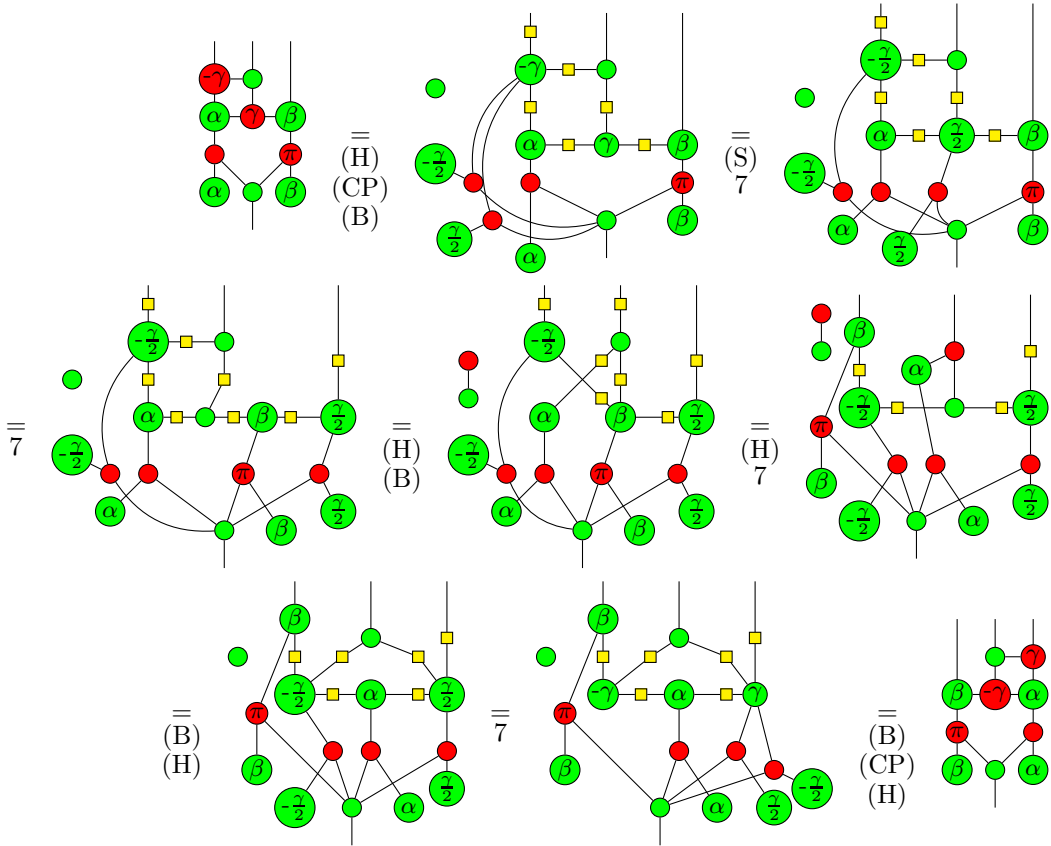
\begin{proof}[Prop. \ref{prop:rule-C}]
\def\fig{C-proof}
\begin{align*}
\input{./figures/\fig/\fig_00.tikz}
\eq{\h\\\bo\\\bt}\input{./figures/\fig/\fig_01.tikz}
\eq{\soo\\\ref{lem:C1-h}}\input{./figures/\fig/\fig_02.tikz}\\
\eq{\ref{lem:C1-h}}\input{./figures/\fig/\fig_03.tikz}
\eq{\h\\\bt}\input{./figures/\fig/\fig_04.tikz}
\eq{\h\\\ref{lem:C1-h}}\input{./figures/\fig/\fig_05.tikz}\\
\eq{\bt\\\h}\input{./figures/\fig/\fig_06.tikz}
\eq{\ref{lem:C1-h}}\input{./figures/\fig/\fig_07.tikz}
\eq{\bt\\\bo\\\h}\input{./figures/\fig/\fig_08.tikz}
\end{align*}
\end{proof}

\begin{proof}[Prop. \ref{prop:add-axiom}]
The idea of the proof is here again to use the SVD, but this time of a state, the equation being between two states. The diagram on the right hand side has an SVD by direct use of Proposition \ref{prop:state-SVD}. However, the one on the left hand side has a cycle, so we have to work a bit more. Notice that, using Proposition \ref{prop:1-1-diag-SVD}:
\def\fig{add-axiom-LHS-interm-SVD}
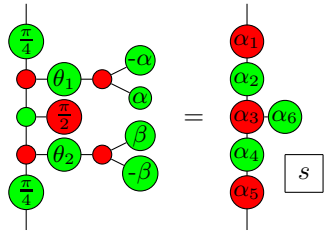
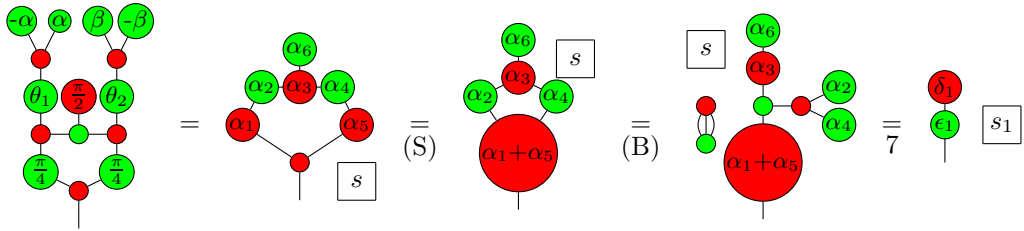
\[\input{./figures/\fig/\fig_00.tikz}\eq{}\input{./figures/\fig/\fig_01.tikz}\]
Hence:
\def\fig{add-axiom-LHS-to-SVD}
\[\input{./figures/\fig/\fig_00.tikz}
\eq{}\input{./figures/\fig/\fig_01.tikz}
\eq{\soo}\input{./figures/\fig/\fig_02.tikz}
\eq{\bt}\input{./figures/\fig/\fig_03.tikz}
\eq{\ref{prop:state-SVD}}\input{./figures/\fig/\fig_04.tikz}\]
From a direct use of Proposition \ref{prop:state-SVD}:
\def\fig{add-axiom-RHS-to-SVD}
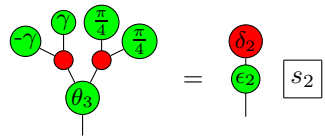
\[\input{./figures/\fig/\fig_00.tikz}
\eq{}\input{./figures/\fig/\fig_01.tikz}\]
Since the initial equation is sound under the constraint $2e^{i\theta_3}\cos{\gamma}=e^{i\theta_1}\cos{\alpha}+e^{i\theta_2}\cos{\beta}$, we have:
\[\def\fig{add-axiom-LHS-to-SVD}\interp{\input{./figures/\fig/\fig_04.tikz}}
\eq{}\def\fig{add-axiom-RHS-to-SVD}\interp{\input{./figures/\fig/\fig_01.tikz}}\]
By Proposition \ref{prop:state-SVD-uniqueness}, either $\delta_1=\delta_2=0$, or $\delta_1=\delta_2=\pi$, or $\delta_1=\delta_2\neq 0\bmod \pi$ and $\epsilon_1=\epsilon_2$. Notice however that in our case, $\delta_1$ and $\delta_2$ cannot be $\pi$ because $\def\fig{add-axiom-RHS-to-SVD}\interp{\input{./figures/\fig/\fig_00.tikz}}=(1+i)\begin{pmatrix}1\\\sqrt{2}e^{i\theta_3}\cos{\gamma}\end{pmatrix}$ and $\interp{
\input{./figures/SVD-state-with-pi.tikz}
}=\begin{pmatrix}0\\\sqrt{2}se^{i\epsilon}\end{pmatrix}$.\\
If $\delta_1=\delta_2=0$, then $\epsilon_1$ and $\epsilon_2$ can be discarded: 
\input{./figures/SVD-state-with-0.tikz}
. Hence, in any case, it only remains to prove that the two scalar diagrams $s_1$ and $s_2$ can be transformed into one another.
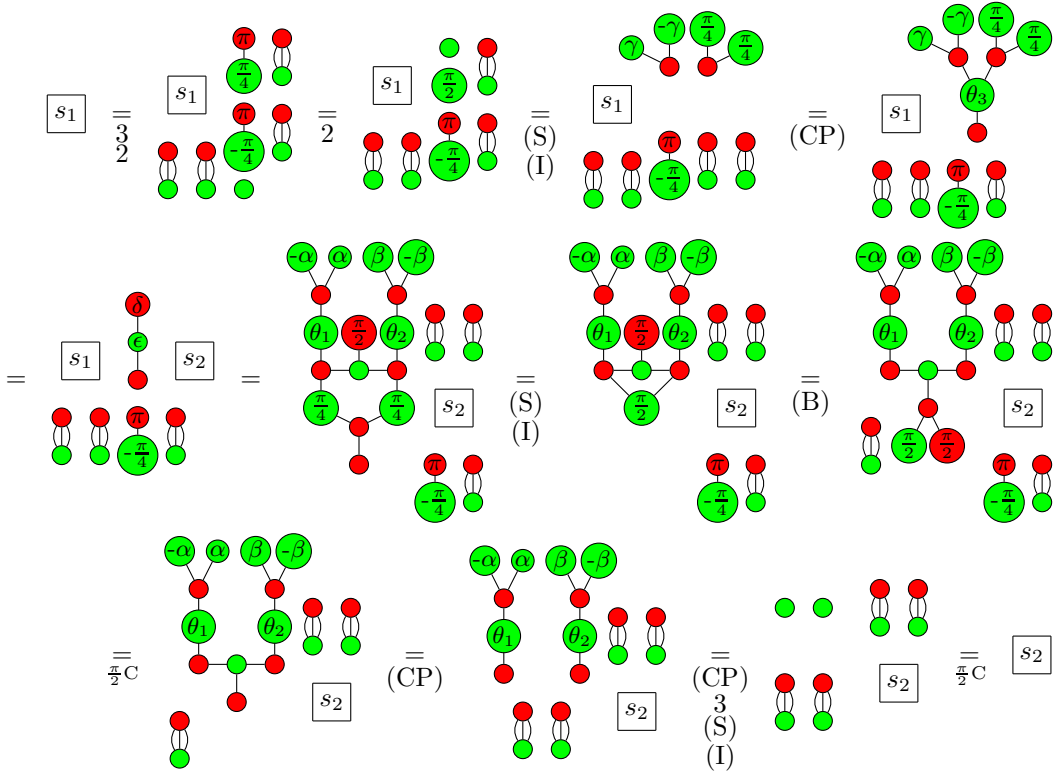
\def\fig{add-axiom-scalars}
\begin{align*}
\input{./figures/\fig/\fig_00.tikz}
\eq{\ref{lem:bicolor-0-alpha}\\\ref{lem:multiplying-phases}}\input{./figures/\fig/\fig_01.tikz}
\eq{\ref{rem:gn-pi_2-0-0-equals-sqrt2-exp-pi_4}}\input{./figures/\fig/\fig_02.tikz}
\eq{\soo\\\stt}\input{./figures/\fig/\fig_03.tikz}
\eq{\bo}\input{./figures/\fig/\fig_04.tikz}\\
\eq{}\input{./figures/\fig/\fig_05.tikz}
\eq{}\input{./figures/\fig/\fig_06.tikz}
\eq{\soo\\\stt}\input{./figures/\fig/\fig_07.tikz}
\eq{\bt}\input{./figures/\fig/\fig_08.tikz}\\
\eq{\comp2}\input{./figures/\fig/\fig_09.tikz}
\eq{\bo}\input{./figures/\fig/\fig_10.tikz}
\eq{\bo\\\ref{lem:bicolor-0-alpha}\\\soo\\\stt}\input{./figures/\fig/\fig_11.tikz}
\eq{\comp2}\input{./figures/\fig/\fig_12.tikz}
\end{align*}
\end{proof}

\begin{proof}[Thm.~\ref{thm:equivalence}]
\phantomsection\label{prf:equivalence}
We are going to prove here that all the equations in Figure \ref{fig:ZX_rules} can be recovered from the ones in \ref{fig:ZX_rules2}.
First of all, we try to recover a complete axiomatisation for Clifford. To do so, we simply need to show that the Hadamard gate can be decomposed.
\def\fig{hopf}
\begin{align}
\label{eq:hopf-EU2}
\input{./figures/\fig/\fig_00.tikz}
\eq{}\input{./figures/\fig/\fig_01.tikz}
\eq{\stt\\\soo}\input{./figures/\fig/\fig_02.tikz}
\eq{\bt}\input{./figures/\fig/\fig_03.tikz}
\eq{\bo}\input{./figures/\fig/\fig_04.tikz}
\eq{\soo\\\stt}\input{./figures/\fig/\fig_05.tikz}
\end{align}
\def\fig{hadamard-involution}
\begin{align}
\label{eq:hadamard-involution}
\input{./figures/\fig/\fig_00.tikz}
\eq{\sttg}\input{./figures/\fig/\fig_01.tikz}
\eq{\h}\input{./figures/\fig/\fig_02.tikz}
\eq{\sttr}\input{./figures/\fig/\fig_03.tikz}
\end{align}
\def\fig{had-decomp-from-EU2}
\begin{align}
\label{eq:had-decomp-from-EU2}
\input{./figures/\fig/\fig_00.tikz}
\eq{\stt}\input{./figures/\fig/\fig_01.tikz}
\eq{\eulerp}\input{./figures/\fig/\fig_02.tikz}
\eq{\h\\(\ref{eq:hadamard-involution})}\input{./figures/\fig/\fig_03.tikz}
\end{align}
\def\fig{rn-0-1-pi_2-to-gn-EU2}
\begin{align}
\label{eq:rn-0-1-pi_2-to-gn-EU2}
\input{./figures/\fig/\fig_00.tikz}
\eq{\h}\input{./figures/\fig/\fig_01.tikz}
\eq{(\ref{eq:had-decomp-from-EU2})\\\soo}\input{./figures/\fig/\fig_02.tikz}
\eq{\iv\\\bo}\input{./figures/\fig/\fig_03.tikz}
\eq{\iv}\input{./figures/\fig/\fig_04.tikz}
\end{align}
\def\fig{had-decomp-scalar-free-from-EU2}
\begin{align}
\label{eq:scalar-free-h-decomp}
\input{./figures/\fig/\fig_00.tikz}
\eq{(\ref{eq:had-decomp-from-EU2})}\input{./figures/\fig/\fig_01.tikz}
\eq{\soo\\\h}\input{./figures/\fig/\fig_02.tikz}
\eq{(\ref{eq:rn-0-1-pi_2-to-gn-EU2})}\input{./figures/\fig/\fig_03.tikz}
\eq{\h}\input{./figures/\fig/\fig_04.tikz}
\end{align}
We have recovered a complete axiomatisation for Clifford. We now have access to all the lemmas deriving from this axiomatisation. The next step is to prove the equation \e is derivable.
\def\fig{pi-commutation-from-EU2}
\begin{align}
\label{eq:pi-commutation-from-EU2}
\input{./figures/\fig/\fig_00.tikz}
\eq{\h}\input{./figures/\fig/\fig_01.tikz}
\eq{\eulerp}\input{./figures/\fig/\fig_02.tikz}
\eq{\soo\\(\ref{eq:had-decomp-from-EU2})}\input{./figures/\fig/\fig_03.tikz}
\eq{\h}\input{./figures/\fig/\fig_04.tikz}
\end{align}
\def\fig{gn-0-0-pi_2-is-sqrt2-exp-i-omega-EU2}
\begin{align}
\label{eq:gn-0-0-pi_2-is-sqrt2-exp-i-omega-EU2}
\input{./figures/\fig/\fig_00.tikz}
\eq{\h}\input{./figures/\fig/\fig_01.tikz}
\eq{\bo\\\iv}\input{./figures/\fig/\fig_02.tikz}
\eq{(\ref{eq:had-decomp-from-EU2})}\input{./figures/\fig/\fig_03.tikz}
\eq{\h\\\iv}\input{./figures/\fig/\fig_04.tikz}
\end{align}
\def\fig{supplementarity-from-EU2}
\begin{align*}
\input{./figures/\fig/\fig_00.tikz}
\eq{\stt\\\soo}\input{./figures/\fig/\fig_01.tikz}
\eq{\bt}\input{./figures/\fig/\fig_02.tikz}
\eq{(\ref{eq:scalar-free-h-decomp})}\input{./figures/\fig/\fig_03.tikz}\\
\eq{\eulerp}\input{./figures/\fig/\fig_04.tikz}
\eq{\soo\\(\ref{eq:hopf-EU2})}\input{./figures/\fig/\fig_05.tikz}
\eq{\bo\\\iv}\input{./figures/\fig/\fig_06.tikz}
\end{align*}
\begin{align*}
\input{./figures/\fig/\fig_07.tikz}
\eq{}\input{./figures/\fig/\fig_08.tikz}
\eq{}\input{./figures/\fig/\fig_09.tikz}
\eq{}\input{./figures/\fig/\fig_10.tikz}
\end{align*}
\begin{align}
\label{eq:supp-from-EU2}
\input{./figures/\fig/\fig_00.tikz}
\eq{}\input{./figures/\fig/\fig_11.tikz}
\eq{}\input{./figures/\fig/\fig_12.tikz}
\end{align}
\def\fig{rule-E-from-EU2}
\begin{align*}
\input{./figures/\fig/\fig_00.tikz}
\eq{\h}\input{./figures/\fig/\fig_01.tikz}
\eq{(\ref{eq:scalar-free-h-decomp})}\input{./figures/\fig/\fig_02.tikz}
\eq{(\ref{eq:pi-commutation-from-EU2})}\input{./figures/\fig/\fig_03.tikz}
\eq{(\ref{eq:supp-from-EU2})}\input{./figures/\fig/\fig_04.tikz}\\
\eq{(\ref{eq:hopf-EU2})}\input{./figures/\fig/\fig_05.tikz}
\eq{(\ref{eq:gn-0-0-pi_2-is-sqrt2-exp-i-omega-EU2})}\input{./figures/\fig/\fig_06.tikz}
\eq{\iv}\input{./figures/\fig/\fig_07.tikz}
\end{align*}
It now remains to prove the rule \euler can be derived.
We decompose the left hand side diagram as such:
\def\fig{euler-from-EU2}
\begin{align*}
\input{./figures/\fig/\fig_00.tikz}
\eq{}\input{./figures/\fig/\fig_01.tikz}
\eq{}\input{./figures/\fig/\fig_02.tikz}
\eq{}\input{./figures/\fig/\fig_03.tikz}
\end{align*}
where $x$ is considered as a variable, and hence, all the computed angles depend on it, while the angles $\alpha_i$ are fixed. We want to find $x_0$ such that $\beta_3(x_0)+\gamma_1(x_0)=0\bmod \pi$. Let the functions $f$ and $g$ be defined as:
\begin{align*}
&f:x\mapsto \atan{\frac{\tan{\alpha_1}\cos{x}+\tan{\alpha_3}\cos{\alpha_2-x}}{1-\tan{\alpha_1}\cos{x}\tan{\alpha_3}\cos{\alpha_2-x}}}\\
&g:x\mapsto \tan{\alpha_1}\cos{x}+\tan{\alpha_3}\cos{\alpha_2-x}
\end{align*}
Notice that $$g\left(-\frac{\pi}{2}\right) = \tan{\alpha_3}\cos{\alpha_2+\frac{\pi}{2}}\qquad\text{and}\qquad g\left(\frac{\pi}{2}\right) = \tan{\alpha_3}\cos{\alpha_2-\frac{\pi}{2}}$$
Hence, $g\left(-\frac{\pi}{2}\right)g\left(\frac{\pi}{2}\right)\leq 0$. Since $g$ is continuous, by the intermediate value theorem, there exists $x_0\in [\frac{-\pi}{2},\frac{\pi}{2}]$ such that $g(x_0)=0$. Notice now that $f(x_0)=\atan{\frac{0}{1+\tan{\alpha_1}^2\cos{\alpha_2-x_0}^2}} = 0$. Also, it can be computed that $f=\beta_3+\gamma_1\bmod \pi$. Hence, $\beta_3(x_0)+\gamma_1(x_0)=0\bmod \pi$ i.e. $\beta_3(x_0)+\gamma_1(x_0)=n \pi$. Hence, denoting $\beta_i\leftarrow \beta_i(x_0)$ and $\gamma_i\leftarrow \gamma_i(x_0)$:
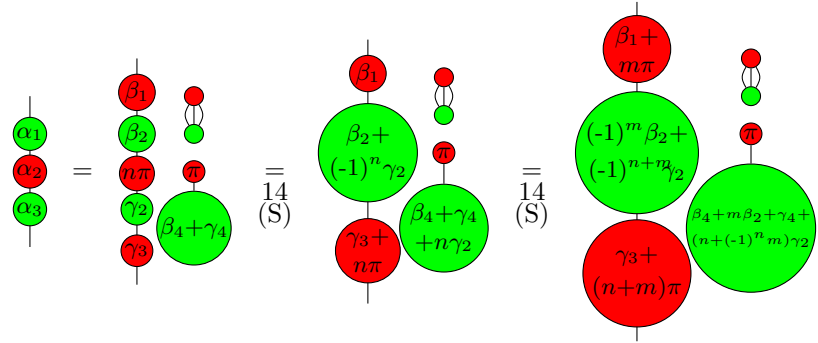
\def\fig{euler-from-EU2}
\begin{align*}
\input{./figures/\fig/\fig_00.tikz}
\eq{}\input{./figures/\fig/\fig_04.tikz}
\eq{\ref{eq:pi-commutation-from-EU2}\\\soo}\input{./figures/\fig/\fig_05.tikz}
\eq{\ref{eq:pi-commutation-from-EU2}\\\soo}\input{./figures/\fig/\fig_06.tikz}
\end{align*}
Since, thank to Proposition \ref{prop:unitary-uniqueness}, the unitary representation is unique if $\beta_1+m\pi\in[0,\pi[$ ($m$ has been chosen for this purpose), then the previous diagram is provably equivalent to the one resulting directly from \euler.
\end{proof}
\end{document}

%% file: figures/Hadamard.tikz
\begin{tikzpicture}
	\begin{pgfonlayer}{nodelayer}
		\node [style={H box}] (0) at (0, 0) {};
		\node [style=none] (1) at (0, 0.5) {};
		\node [style=none] (2) at (0, -0.5) {};
	\end{pgfonlayer}
	\begin{pgfonlayer}{edgelayer}
		\draw (2.center) to (1.center);
	\end{pgfonlayer}
\end{tikzpicture}

%% file: figures/single-line.tikz
\begin{tikzpicture}
	\begin{pgfonlayer}{nodelayer}
		\node [style=none] (0) at (0, 0.2499999) {};
		\node [style=none] (1) at (0, -0.2499999) {};
	\end{pgfonlayer}
	\begin{pgfonlayer}{edgelayer}
		\draw (0.center) to (1.center);
	\end{pgfonlayer}
\end{tikzpicture}

%% file: figures/cup.tikz
\begin{tikzpicture}
	\begin{pgfonlayer}{nodelayer}
		\node [style=none] (0) at (-0.2500001, 0.2500001) {};
		\node [style=none] (1) at (0.2500001, 0.2500001) {};
	\end{pgfonlayer}
	\begin{pgfonlayer}{edgelayer}
		\draw [bend right=90, looseness=1.75] (0.center) to (1.center);
	\end{pgfonlayer}
\end{tikzpicture}

%% file: figures/caps.tikz
\begin{tikzpicture}
	\begin{pgfonlayer}{nodelayer}
		\node [style=none] (a0) at (-0.2500001, -0) {};
		\node [style=none] (a1) at (0.2500001, -0) {};
		\node [style=none] (a2) at (0, 0.25) {};
	\end{pgfonlayer}
	\begin{pgfonlayer}{edgelayer}
		\draw [bend left=90, looseness=1.75] (a0.center) to (a1.center);
	\end{pgfonlayer}
\end{tikzpicture}

%% file: figures/generic-state-1.tikz
\begin{tikzpicture}
	\begin{pgfonlayer}{nodelayer}
		\node [style=rn] (0) at (0, 0.5) {$\alpha_1$};
		\node [style=gn] (1) at (0, -0) {$\beta_1$};
		\node [style=none] (2) at (0, -0.5) {};
	\end{pgfonlayer}
	\begin{pgfonlayer}{edgelayer}
		\draw [style=none] (2.center) to (0);
	\end{pgfonlayer}
\end{tikzpicture}

%% file: figures/generic-state-2.tikz
\begin{tikzpicture}
	\begin{pgfonlayer}{nodelayer}
		\node [style=rn] (0) at (0, 0.5) {$\alpha_2$};
		\node [style=gn] (1) at (0, -0) {$\beta_2$};
		\node [style=none] (2) at (0, -0.5) {};
	\end{pgfonlayer}
	\begin{pgfonlayer}{edgelayer}
		\draw [style=none] (2.center) to (0);
	\end{pgfonlayer}
\end{tikzpicture}

%% file: figures/C1-proof-branch-to-remove.tikz
\begin{tikzpicture}
	\begin{pgfonlayer}{nodelayer}
		\node [style=gn] (0) at (0.5, -0.25) {$\beta_2$};
		\node [style=none] (1) at (0, -0.5) {};
		\node [style=rn] (2) at (0, 0) {};
		\node [style=none] (3) at (0, 0.5) {};
	\end{pgfonlayer}
	\begin{pgfonlayer}{edgelayer}
		\draw [style=none] (0) to (2);
		\draw (3.center) to (1.center);
	\end{pgfonlayer}
\end{tikzpicture}